\documentclass[12pt]{article}

\usepackage{textcomp}
\usepackage{chetnew}
\usepackage{tikz}

\newcommand{\CH}{\mathcal{H}}
\newcommand{\CC}{\mathcal{C}}
\newcommand{\CO}{\mathcal{O}}
\newcommand{\CT}{\mathcal{T}}

\newcommand{\CN}{\mathcal{N}}
\newcommand{\CS}{\mathcal{S}}
\newcommand{\CM}{\mathcal{M}}

\preprint{RU-NHETC-2014-20\\ QMUL-PH-14-24}

\title{Argyres-Douglas Theories and $S$-Duality}

\author{Matthew~Buican,$^{1\diamondsuit}$ Simone~Giacomelli,$^{2\clubsuit}$ Takahiro~Nishinaka,$^{1\heartsuit}$ and Constantinos~Papageorgakis$^{3\spadesuit}$
}

\affiliation{$^1$ NHETC and Department of Physics and Astronomy \\ Rutgers University, Piscataway, NJ 08854, USA\\$^2$ Universit\'e Libre de Bruxelles and International Solvay Institutes\\ ULB-Campus Plaine CP231 1050 Brussels, Belgium\\ $^3$ CRST and School of Physics and Astronomy\\ Queen Mary University of London, E1 4NS, UK
\emails{$^{\diamondsuit}$buican@physics.rutgers.edu,$^{\clubsuit}$simone.giacomelli@ulb.ac.be,$^{\heartsuit}$nishinaka@physics.rutgers.edu, $^{\spadesuit}$c.papageorgakis@qmul.ac.uk}}

\abstract{We generalize $S$-duality to $\CN=2$ superconformal field theories (SCFTs) with Coulomb branch operators of non-integer scaling dimension. As simple examples, we find minimal generalizations of the $S$-dualities discovered in $SU(2)$ gauge theory with four fundamental flavors by Seiberg and Witten and in $SU(3)$ gauge theory with six fundamental flavors by Argyres and Seiberg. Our constructions start by weakly gauging diagonal $SU(2)$ and $SU(3)$ flavor symmetry subgroups of two copies of a particular rank-one Argyres-Douglas theory (along with sufficient numbers of hypermultiplets to guarantee conformality of the gauging). As we explore the resulting conformal manifold of the $SU(2)$ SCFT, we find an action of $S$-duality on the parameters of the theory that is reminiscent of ${\rm Spin}(8)$ triality. On the other hand, as we explore the conformal manifold of the $SU(3)$ theory, we find that an exotic rank-two SCFT emerges in a dual $SU(2)$ description.}
\date{November 2014}

\begin{document}
\maketitle

\toc

\newsec{Introduction and Summary}[Intro]
$\CN=2$ superconformal field theories (SCFTs) often have exactly marginal deformations that preserve $\CN=2$ supersymmetry (SUSY). Such deformations are descendants of dimension two operators that we can add to the prepotential
\eqn{
\delta S=\int d^2\theta_1d^2\theta_2\ \lambda^i\CO_i+{\rm h.c.}~,
}[marginal]
where the integration is taken over the $\CN=2$ chiral Grassmann parameters. The $\lambda^i$ parameterize spaces commonly referred to as \lq\lq conformal manifolds."\foot{On general grounds, such conformal manifolds are K\"ahler \rcite{Asnin:2009xx}.} Simple examples of theories with exactly marginal couplings include all the Lagrangian theories (e.g., $\CN=4$ Super Yang-Mills, $SU(N_c)$ gauge theory with $N_f=2N_c$ fundamental flavors, etc.).

Often, the $\lambda^i$ can be interpreted as gauge couplings with vanishing beta functions: the resulting conformal manifolds have cusps where perturbative gauge fields emerge and couple various isolated theories.\foot{They can also couple sectors with co-dimension one or higher conformal manifolds. However, we can often continue this process iteratively until we have a collection of isolated theories.} By definition, the isolated SCFT sectors do not have their own exactly marginal deformations. Instead, they have global symmetries that are weakly gauged.

The simplest isolated SCFTs we can consider gauging are just collections of free hypermultiplets. For example, taking a collection of eight hypermultiplets and gauging an $SU(2)\subset Sp(8)$ flavor subgroup, we construct the $SU(2)$ theory with $N_f=4$ and $SO(8)$ flavor symmetry. As we vary the resulting exactly marginal coupling, $\tau={\theta\over\pi}+{8\pi i\over g^2}$, the theory becomes strongly coupled. However, if we tune the coupling appropriately, a new weakly coupled $S$-dual description emerges at another cusp\rcite{Seiberg:1994aj} which looks like the original theory up to an $S_3$ triality outer automorphism of the flavor ${\rm Spin}(8)$. The duality group in this case is $SL(2, {\bf Z})$, and this construction extends the notion of $\CN=4$ duality \rcite{Montonen:1977sn,Goddard:1976qe,Witten:1978mh,Osborn:1979tq} to an $\CN=2$ theory.

More generally, if one starts from a Lagrangian theory and tunes the gauge coupling to another cusp, one often finds that a new isolated interacting SCFT emerges. For example, in \rcite{Argyres:2007cn}, Argyres and Seiberg found that, by starting with the weakly coupled $SU(3)$ gauge theory with six fundamental flavors and varying the gauge coupling, a new cusp emerges with an $S$-dual description in which the Minahan-Nemeschansky (MN) theory with $E_6$ global symmetry \rcite{Minahan:1996fg} is weakly coupled to a doublet of $SU(2)$ via an $SU(2)\subset E_6$ gauging. This type of duality has been generalized by Gaiotto \rcite{Gaiotto:2009we} and many other authors (see, e.g., \rcite{Argyres:2007tq} and \rcite{Chacaltana:2010ks}).

All other examples of $S$-duality discussed in the literature essentially share the general characteristics of the above two cases, but with varying numbers of cusps and isolated sectors of varying ranks (i.e., varying dimensions of their Coulomb branches). In particular, all the instances of $S$-duality that we are aware of involve $\CN=2$ scalar chiral primaries (we mean operators annihilated by all the anti-chiral Poincar\'e supercharges; these operators are often called \lq\lq Coulomb branch" operators) of integer dimension.

In this paper, we will generalize $S$-duality to theories with non-integer scaling dimension Coulomb branch primaries. Since Lagrangian theories have only integer dimension $\CN=2$ chiral operators, our theories of interest are never completely weakly coupled. Instead, we will find various cusps where weakly coupled gauge fields emerge and couple various isolated strongly coupled sectors that are related to each other in interesting ways.\foot{Note that there can be conformal manifolds with only integer dimension Coulomb branch operators that do not have a Lagrangian limit because they have some exceptional flavor symmetry (for example, one can gauge an $SU(3)$ subgroup of the flavor symmetry of the $E_8$ SCFT as in \rcite{Argyres:2007tq}).}

The original examples of theories with non-integer dimension chiral operators were discovered as special points in the Coulomb branch of $SU(3)$ super Yang-Mills by Argyres and Douglas \rcite{Argyres:1995jj} and in $SU(2)$ SQCD with $N_f=1,2,3$ flavors in \rcite{Argyres:1995xn} (the $N_f=1$ SCFT is the same as the one in \rcite{Argyres:1995jj}). Following the notation of \rcite{Xie:2013jc}, we will refer to these theories as the $I_{2,3}$, $I_{2,4}$, and $I_{3,3}$ SCFTs respectively.\footnote{These SCFTs also go under many different names. For example, they are sometimes referred to as the $(A_1, A_2)$, $(A_1, A_3)$, and $(A_1, D_4)$ theories \rcite{Cecotti:2010fi} (this notation arises from the fact that the BPS quivers of these theories are the products of the corresponding ADE Dynkin diagrams).} These theories are believed to be the only rank-one SCFTs with non-integer dimension $\CN=2$ chiral operators.\foot{More precisely, Kodaira's classification of elliptic fibrations over the complex plane \rcite{Kodaira:1963xx} implies that the only consistent non-integer scaling dimensions of $\CN=2$ Coulomb branch generators describing a rank one theory are $6/5$, $4/3$, or $3/2$. These scaling dimensions are realized, respectively, by the $I_{2,3}$, $I_{2,4}$, and $I_{3,3}$ theories \cite{Argyres:1995xn}. While it is not inconceivable that other inequivalent theories have the same spectrum, no such theories have been found to date.} Of course, there are also many higher-rank Argyres-Douglas (AD) theories (e.g., see the review in \rcite{Giacomelli:2013tia}).

Although the above AD theories are isolated, they typically inherit some flavor symmetry from the UV gauge theories in which they are embedded. For example, the $I_{2,4}$ and $I_{3,3}$ theories have $SU(2)$ and $SU(3)$ flavor symmetry respectively (the $I_{2,3}$ theory has no flavor symmetry). Therefore, we can try gauging the flavor symmetries of the $I_{2,4}$ or $I_{3,3}$ theories in an exactly marginal fashion (adding additional sectors charged under a diagonal combination of flavor symmetries as necessary), studying the resulting conformal manifold, and finding the various $S$-dual frames.

To that end, in the first part of this paper, we will study a particular rank three theory, which we denote as $\CT_{2,{3\over2},{3\over2}}$. This theory consists of $SU(2)$ gauge fields coupled to two $I_{3,3}$ theories and a doublet of hypermultiplets. As a result, $\CT_{2,{3\over2},{3\over2}}$ has one marginal coupling. We will see that this marginal coupling parameterizes a conformal manifold with three $S$-dual cusps, and that, at each of these cusps, $SU(2)$ gauge fields emerge and couple two $I_{3,3}$ theories and a doublet of hypermultiplets (with the parameters of the theory mixed in interesting ways). After appropriately taking into account the mixing of the different parameters, we will find an analog of the triality discussed in \rcite{Seiberg:1994aj}. Furthermore, subject to some assumptions, we will prove that the $\CT_{2,{3\over2},{3\over2}}$ theory is the minimal (i.e., lowest-rank) theory with non-integer dimensional Coulomb branch operators that has a marginal gauge coupling and exhibits $S$-duality. As such, our discussion of the $\CT_{2,{3\over2},{3\over2}}$ SCFT represents the minimal generalization of Seiberg and Witten's analysis of $SU(2)$ with $N_f=4$ \rcite{Seiberg:1994aj}.

In the second part of the paper, we look for the lowest-rank generalization of Argyres-Seiberg duality. We will argue that such a generalization is given by a rank four theory we call $\CT_{3,2,{3\over2},{3\over2}}$ (note that there may be other rank four generalizations). It consists of an $SU(3)$ gauge theory coupled to two $I_{3,3}$ theories with three fundamental flavors of $SU(3)$ (i.e., we replace half the fundamentals of the $N_f=6$ $SU(3)$ gauge theory with $I_{3,3}$ sectors). Interestingly, at another cusp, $\CT_{3,2,{3\over2},{3\over2}}$ has an $SU(2)$ gauge theory realization in which the gauge group is coupled to a single $I_{3,3}$ theory and a more exotic theory of rank two with Coulomb branch spectrum\foot{Throughout this paper, we take the term \lq\lq spectrum" in this context to mean the spectrum of {\it generators} of the Coulomb branch chiral ring.} $\left\{3,{3\over2}\right\}$ that we will call $\CT_{3,{3\over2}}$ (this theory plays the role of the $E_6$ SCFT in our duality). The latter has a $G_{\CT_{3,{3\over2}}}\supset SU(3)\times SU(2)$ flavor symmetry, of which we gauge the $SU(2)$ factor. $\CT_{3,{3\over2}}$ has not been explicitly discussed in the literature (although it appears implicitly in the classification of \rcite{Xie:2012hs,Xie:2013jc}), and our analysis will elucidate some of its interesting properties. For example, our results imply that the $SU(2)\subset G_{\CT_{3,{3\over2}}}$ flavor symmetry does not suffer from Witten's anomaly \rcite{Witten:1982fp}. Moreover, it follows from our analysis that the $SU(2)$ and $SU(3)$ flavor central charges are
\eqn{
k_{SU(2)}^{\CT_{3,{3\over2}}}=5~, \ \ \ k_{SU(3)}^{\CT_{3,{3\over2}}}=6~.
}[ASkSUii]
The result for $k_{SU(2)}^{\CT_{3,{3\over2}}}$ is somewhat unconventional, since it does not follow from the usual rule of thumb for relating flavor central charges to (in our normalization) twice the scaling dimension of some Coulomb branch generator in the theory; indeed, the $\CT_{3,{3\over2}}$ SCFT has no Coulomb branch generator of dimension-${5\over 2}$. Also, using the results of \rcite{Buican:2014qla}, we can immediately conclude that since the $I_{3,3}$ theory does not have exotic $\CN=2$ chiral primaries, neither does the $\CT_{3, {3\over2}}$ theory.

The rest of this paper is organized as follows: In Section~\ref{strategy} we describe the tools that let us identify the AD building blocks in the various S-dual frames. In Section~\ref{SW} we give the details of the rank three example generalizing the $S$-duality of $SU(2)$ with $N_f=4$, while in Section~\ref{AS} we discuss the rank four generalization of Argyres-Seiberg duality. We briefly conclude in Section~\ref{conclusions}. In Appendix~\ref{AppA}, we sketch out the Hitchin system derivation of the various Seiberg-Witten curves we use in the main part of the paper. Appendix~\ref{app:equiv} exhibits the equivalence of the $(III_{3,3}^{3\times[2,1]},F)$ theory to $I_{3,3}$ plus a triplet of hypermultiplets. Finally, in Appendix~\ref{AppC}, we give an independent derivation of the $\CT_{2,{3\over2},{3\over2}}$ and $\CT_{3,2,{3\over2},{3\over2}}$ curves.

\newsec{The Strategy}[strategy]

The idea of using isolated sectors to construct conformal manifolds of $\CN = 2$ SCFTs by weakly gauging flavor symmetry subgroups is rather general. In order to make sense of the vast set of possible building blocks and the $S$-dual cusps that can emerge, we should find some simple, universal, and invariant characterizations of the physics on an $\CN=2$ conformal manifold, $\CM$. For example, we can study:
\begin{itemize}

\item[{\bf(i)}] The $a$ and $c$ conformal anomalies.

\item[{\bf(ii)}] The set of flavor symmetries (in our conventions, these are symmetries commuting with the $\CN=2$ superconformal algebra and not related by supersymmetry to higher-spin symmetries), $G=\prod_iG_i$, and the corresponding flavor central charges, $k_i$.

\item[{\bf(iii)}] The spectrum, $\CS$, of Coulomb branch operators.

\end{itemize}
These quantities do not change as we travel along $\CM$.\foot{The $a$ and $c$ central charges are invariant under exactly marginal deformations by the usual anomaly matching arguments (conformal symmetry is unbroken as we move along $\CM$). The flavor symmetries are also invariant (at the cusps, where weakly coupled gauge fields emerge, we also have emergent flavor symmetries; however, these symmetries are arbitrarily weakly gauged) since the exactly marginal primaries, $\CO_i$, are uncharged under the flavor symmetries (this follows, e.g., from the analysis of the $\CO_i\CO_{\bar j}^{\dagger}$ OPE in \rcite{Buican:2013ica}). As a result, by anomaly matching arguments, the $k_i$ are constant on $\CM$. The invariance of the $\CN=2$ chiral spectrum follows from \rcite{Dolan:2002zh}, which shows that the number of such operators cannot change as we traverse the conformal manifold, and from \rcite{Papadodimas:2009eu}, which shows that the dimensions of these operators do not change either. Note that this reasoning applies also to the \lq\lq exotic" higher-spin $\CN=2$ chiral primaries considered in \rcite{Buican:2014qla}.}

As we go between different cusps of the conformal manifold, the various quantities in {\bf(i)}, {\bf(ii)}, and {\bf(iii)} are ``partitioned'' among the different emergent sectors. One interesting aspect of the Argyres-Seiberg-like dualities is that, unlike in the case of $SU(2)$ gauge theory with $N_f=4$, these quantities are generally distributed differently at the different cusps. For example, in the case of \rcite{Argyres:2007cn}, at the $SU(3)$ cusp we have
\eqna{
a&={5\over3}+6\cdot{1\over8}={29\over12}~, \ \ \ c={4\over3}+6\cdot{1\over4}={17\over6}\\ G&=SU(6)\times U(1)~, \ \ \ k_{SU(6)}=0+6=6~,\ \ \ k_{U(1)}=0+6\cdot 6=36~, \\ \CS&=\left\{2,3\right\}\oplus\emptyset~.
}[SUthreecusp]
The first contributions in $a$ and $c$ come from the $SU(3)$ gauge sector, while the remaining contributions come from the six flavors (this partition reflects the fact that there are seven corresponding $\CN=2$ stress tensor multiplets and hence seven different $\CN=2$ sectors).\foot{In our conventions, $a={3\over32}\left(3{\rm Tr}\tilde R^3-{\rm Tr}\tilde R\right)$ and $c={3\over32}\left(3{\rm Tr}\tilde R^3-{5\over3}{\rm Tr}\tilde R\right)$, where $\tilde R={1\over3}R_{\CN=2}+{4\over3}I_3$ is the $\CN=1\subset\CN=2$ superconformal $R$ charge, $R_{\CN=2}$ is the $\CN=2$ superconformal $U(1)_R\subset U(1)_R\times SU(2)_R$ charge, and $I_3$ is the Cartan of $SU(2)_R$ (a free $\CN=2$ $U(1)$ vector multiplet scalar primary has $I_3=0$ and $R_{\CN=2}=2$).} Finally, the flavor symmetry comes from the hypermultiplets, and the gauge sector gives all the contributions to $\CS$ (the elements of $\CS$ are the scaling dimensions of the generators of the $\CN=2$ chiral ring---in this case the Casimirs of $SU(3)$). On the other hand, at the $SU(2)$ cusp, we find three distinct $\CN=2$ sectors (with three independent $\CN=2$ stress tensor multiplets)
\eqna{
a&={5\over8}+{41\over24}+{1\over12}={29\over12}~, \ \ \ c={1\over2}+{13\over6}+{1\over6}={17\over6}\\ G&=SU(6)\times U(1)~, \ \ \ k_{SU(6)}=0+6+0=6~,\ \ \ k_{U(1)}=0+0+2\cdot18=36~, \\ \CS&=\left\{2\right\}\oplus\left\{3\right\}\oplus\emptyset~.
}[SUtwocusp]
The first contributions in the above partitions are from the gauge sector, the second contributions come from the MN theory, which has rank one (its Coulomb branch chiral ring has a single generator of dimension three), and the third contributions come from the doublet of hypermultiplets.

Now let us turn to  theories with non-integer dimensional operators. In the case of the $\CT_{2,{3\over2},{3\over2}}$ theory mentioned in the introduction, we have
\eqna{
&a_{\CT_{2,{3\over2},{3\over2}}}={15\over24}+2\cdot{7\over12}+{1\over12}={15\over8}~, \ \ \ c_{\CT_{2,{3\over2},{3\over2}}}={1\over2}+2\cdot{2\over3}+{1\over6}=2~,\\ & G_{\CT_{2,{3\over2},{3\over2}}}=U(1)^3~, \ \ \ k_{U(1)^3}=(3,3,3)~, \\ & \CS_{\CT_{2,{3\over2},{3\over2}}}=\left\{2\right\}\oplus\left\{3\over2\right\}\oplus\left\{3\over2\right\}\oplus\emptyset~,
}[Trkiiidata]
where the first contributions are from the $SU(2)$ gauge sector, the second and third contributions are from the two $I_{3,3}$ SCFTs, and the final contributions are from the hypermultiplets. The remaining flavor symmetry is $U(1)^3$ since we gauge a diagonal $SU(2)\subset SU(3)\times SU(3)\times Sp(2)$, where the $SU(3)$ factors come from the $I_{3,3}$ sectors and the $Sp(2)$ factor comes from the two hypermultiplets. This gauging is marginal since $k_{SU(2)}=2k_{SU(2)}^{I_{3,3}}+k_{{\bf2}\oplus{\bf2}}=2\cdot 3+2=8$.\foot{Here we use the fact that $SU(3)$ flavor central charge is $k^{I_{3,3}}_{SU(3)}=3$ \rcite{Aharony:2007dj}. Furthermore, we have $k_{SU(2)}^{I_{3,3}}=k_{SU(3)}^{I_{3,3}}$, since the embedding index of $SU(2)\subset SU(3)$ is unity.}

On the other hand, in the case of the $\CT_{3,2,{3\over2},{3\over2}}$ SCFT, we have
\eqna{
&a_{\CT_{3,2,{3\over2},{3\over2}}}={5\over3}+2\cdot{7\over12}+3\cdot{1\over8}={77\over24}~, \ \ \ c_{\CT_{3,2,{3\over2},{3\over2}}}={4\over3}+2\cdot{2\over3}+3\cdot{1\over4}={41\over12}~,\\ & G_{\CT_{3,2,{3\over2},{3\over2}}}=U(3)~, \ \ \ k_{SU(3)}=0+0+3\cdot 2=6~, \ \ \ k_{U(1)}=0+0+3=3~,\\ & \CS_{{\CT_{3,2,{3\over2},{3\over2}}}}=\left\{3, 2\right\}\oplus\left\{3\over2\right\}\oplus\left\{3\over2\right\}\oplus\emptyset~,
}[Trkivdata]
where the first contributions are from the $SU(3)$ gauge sector, the second and third contributions are from the two $I_{3,3}$ SCFTs, and the final contributions are from the hypermultiplets. The flavor symmetry is $U(3)$ since we gauge a diagonal $SU(3)\subset SU(3)\times SU(3)\times Sp(9)$, where the last factor comes from the hypermultiplets. We again have a marginal gauging since $k_{SU(3)}=2k_{SU(3)}^{I_{3,3}}+3k_{{\bf3}\oplus{\bf\bar3}}=2\cdot 3+3\cdot2=12$.

Our strategy for exploring the various cusps of the $\CT_{2,{3\over2},{3\over2}}$ and $\CT_{3,2,{3\over2},{3\over2}}$ conformal manifolds is simple. We first take the data in \Trkiiidata\ and \Trkivdata\ and match it to data for the corresponding theories in the infinite class of AD SCFTs described in \rcite{Xie:2012hs, Xie:2013jc}. In particular, we will argue that
\eqn{
\CT_{2,{3\over2},{3\over2}}=I_{4,4}~, \ \ \ \CT_{3,2,{3\over2},{3\over2}}=III_{6,6}^{3\times[2,2,1,1]}~,
}[match]
where the theories listed on the RHS of \match\ are defined in \rcite{Xie:2012hs, Xie:2013jc}.\foot{Evidence for the first equality in \match\ was presented at the level of the BPS spectra in \rcite{Cecotti:2011rv} (note that the methods in \rcite{Gaiotto:2012db} are also useful for finding the BPS spectrum in this case). We will describe how $S$-duality works in this theory. Note also that, as we explain in more detail below, the superscript \lq\lq$3\times[2,2,1,1]$" in the second equality refers to certain Young tableaux that define the $III_{6,6}^{3\times[2,2,1,1]}$ SCFT.} Using our methods, it is clear that one can explore infinitely many generalizations of the conformal manifolds we will discuss in this text.

In \match, $I_{4,4}$ and $III_{6,6}^{3\times[2,2,1,1]}$ are low-energy theories coming from $M5$-branes wrapping a Riemann sphere with one irregular puncture (they can be thought of as twisted compactifications of the $A_k$ $(2,0)$ theory and are therefore referred to as being of class $\CS$).\foot{In fact, there is some redundancy in this description, and, as we will see, both the $\CT_{2,{3\over2},{3\over2}}$ and $\CT_{3,2,{3\over2},{3\over2}}$ theories can also be realized as the IR description of $M5$ branes wrapping a sphere with one irregular and one regular puncture.} These theories can be succinctly described in terms of Hitchin systems,\foot{See \rcite{Gaiotto:2009hg} for a beautiful account of the relationship between theories of class $\CS$ and Hitchin systems.} and the corresponding Seiberg-Witten (SW) curves come from the spectral covers of these Hitchin systems. Using the resulting curves, we can then explore the various cusps of the conformal manifolds and find new $S$-dual frames. As an alternate derivation, we will also show how to obtain the SW curves directly from certain UV-complete linear quiver theories.

Crucially, the Hitchin systems also give us direct access to the quantities {\bf(i)-(iii)} without the need to fully analyze the SW curves.\foot{One possible exception to this statement might be the set of flavor anomalies.} As a result, we can immediately generate conjectures about different $S$-dual frames and perform some checks on our guesses before verifying them by analyzing the SW curve. Indeed, in the examples below, we will essentially be able to conjecture the $S$-dualities from studying the different ways in which the quantities in {\bf(i)-(iii)} can be partitioned. To confirm these guesses, we then study various limits of the SW curve. 

The reasons we can proceed in this way are as follows:
\begin{itemize}
\item[$\bullet$] The Casimirs of the adjoint Higgs field in the Hitchin system description allow us to find the Coulomb branch spectrum, $\CS=\left\{\Delta_1, \cdots, \Delta_N\right\}$. By the results of \rcite{Shapere:2008zf}, this data also fixes\foot{A condition for using the results in \rcite{Shapere:2008zf} is that our theory has a freely generated Coulomb branch. All the theories we study in this paper satisfy this condition.}
\end{itemize}
\eqn{
2a-c={1\over2}\sum_{i=1}^N\left(\Delta_i-{1\over2}\right)~.
}[ShapTach]
\begin{itemize}
\item[$\bullet$] Using the recipes in \rcite{Xie:2012hs, Xie:2013jc, Nanopoulos:2010bv} (see also the discussion in \rcite{Boalch:XXXXxx} and \rcite{Boalch:YYYYyy}), we can give a Lagrangian description of the three-dimensional mirror of the $S^1$ compactification of our theory, $\CT_{\rm3dm}$. Although this description is not always ``good'' (in the sense that the IR superconformal $R$-symmetry can mix with accidental symmetries), we can unambiguously compute the dimension of the corresponding Coulomb branch, ${\rm dim}\CM_{\CC}^{\rm 3dm}$, and hence $a-c$ via the relation
\end{itemize}
\eqn{
a-c=-{1\over24}{\rm dim}\CM_{\CH}=-{1\over24}{\rm dim}\CM_{\CC}^{\rm 3dm}~.
}[amc]
\begin{itemize}
\item[]We expect \amc\ to hold in all theories that have a genuine Higgs branch (all the superconformal theories of class $\CS$ discussed in \rcite{Xie:2012hs, Xie:2013jc} with non-integer dimension Coulomb branch operators come from genus zero compactifications of the $(2,0)$ theory and therefore have Higgs branches).\foot{The first equality in \amc\ is a natural generalization, to strongly coupled theories with a Higgs branch, of the weakly coupled result that $a-c=-{1\over24}(n_H-n_V)$, where $n_H$ is the number of hypermultiplets and $n_V$ is the number of vector multiplets. The second equality in \amc\ follows from mirror symmetry (in particular, the exchange of Higgs and Coulomb branches under this duality) and the fact that the Higgs branch does not receive quantum corrections as we go to long distances compared to the $S^1$ radius.}
\item[$\bullet$] The three-dimensional mirror often allows us (as long as the IR behavior is under sufficient control) to fix the precise flavor symmetry of the theory via the monopole analysis of \rcite{Gaiotto:2008ak} or, sometimes, from applying mirror symmetry again and reading off the flavor symmetry directly.\foot{As we will discuss, in the case of the $\CT_{3,{3\over2}}$ theory, this analysis is somewhat more subtle.}
\end{itemize}

We should note that from the perspective of the compactification of the $A_k$ $(2,0)$ theory, it may be somewhat surprising that we have an exactly marginal parameter at all. Indeed, in the case of Gaiotto's theories \rcite{Gaiotto:2009we}, marginal parameters in the four-dimensional field theory are identified with complex structure deformations of the Riemann surfaces on which the parent six-dimensional theory is compactified. Clearly, the punctured spheres we consider do not have any complex structure deformations. Instead, it turns out that the exactly marginal deformations in our theories arise from certain dimensionless parameters of the co-dimension two defects used in defining the six-dimensional parent theory.\foot{We thank G.~Moore for a discussion of this point.}

Finally, before we proceed, we should also note that in studying the behavior of our theories at different cusps in the marginal coupling space, we will often find it necessary to renormalize some of our parameters by multiplying them by functions that either vanish or diverge at a given cusp. The reason we do this is simple. We must demand that our parameterization of the Coulomb branch is non-singular so that the BPS masses are finite and non-trivial functions of the Coulomb branch coordinates. Presumably this criterion can be also understood as the necessity of renormalizing the operators whose vevs parameterize the Coulomb branch as we traverse the conformal manifold. In \rcite{Papadodimas:2009eu}, this renormalization was interpreted as the statement that operators can pick up non-trivial phases or mix in interesting ways as we travel along closed loops in the marginal coupling space (i.e., operators transform as sections of certain bundles over the conformal manifold). We will find some evidence for this picture, since our normalizations introduce monodromies in the marginal parameter space.

\newsec{A minimal generalization of Seiberg and Witten's $S$-duality}[SW]
In this section, we will study the $\CT_{2,{3\over2},{3\over2}}$ theory introduced above. In the first subsection, we find the invariant quantities {\bf(i)-(iii)} of the $I_{4,4}$ theory \rcite{Xie:2012hs} and show that they match those of $\CT_{2,{3\over2},{3\over2}}$.\foot{Note that we can also realize our theory in terms of the $(I_{3,3}, S)$ Hitchin system. This system has lower rank than the $I_{4,4}$ Hitchin system, but it also has an additional regular singularity.} We also argue that, subject to some assumptions, the only potential cusps of the $\CT_{2,{3\over2},{3\over2}}$ theory involve an $SU(2)$ gauge sector coupled to two $I_{3,3}$ sectors and a doublet of hypermultiplets (in other words, we argue that there is no emergent rank-two sector with Coulomb branch spectrum $\left\{{3\over2},{3\over2}\right\}$).

We then find further evidence for this picture by analyzing the SW curve of the $I_{4,4}$ theory. Moreover, we find an $S$-duality action on the parameters of the theory that is reminiscent of the ${\rm Spin}(8)$ triality of the $SU(2)$ gauge theory with $N_f=4$. As a result, this discussion represents a simple generalization of Seiberg and Witten's analysis \rcite{Seiberg:1994aj}. In the final subsection, we show how $\CT_{2,{3\over2},{3\over2}}$ can be derived from a UV-complete linear quiver.

Before proceeding to the calculations, let us show that our theory is the simplest (i.e., lowest-rank) example of an $S$-duality with non-integer dimension Coulomb branch operators under certain reasonable assumptions: {\bf(a)} the only rank-zero theories are collections of free hypermultiplets and {\bf(b)}  the only rank-one theories with non-integer scaling dimension primaries are the $I_{2,3}$, $I_{2,4}$, and $I_{3,3}$ theories.\foot{It might be possible to prove assumption {\bf(a)} by generalizing \rcite{Shapere:2008zf} and using the $\CN=2$ version of the arguments presented in \rcite{Zhiboedov:2013opa}.}

Under these assumptions, it follows from the fact that $I_{2,3}$ has no flavor symmetry and the fact that $k_{SU(2)}^{I_{2,4}}={8\over3}$ \rcite{Aharony:2007dj} that the lowest rank theory we can imagine constructing---let us call it $\CT_{\rm rk2}$--- involves an $SU(2)$ gauge theory coupled to one copy of the $I_{3,3}$ theory (via a gauging of the $SU(2)\subset SU(3)$ flavor symmetry) and five hypermultiplets (via a gauging of $SU(2)\subset Sp(5)$) so that $k_{SU(2)}=k_{SU(2)}^{I_{3,3}}+5k_{{\bf2}}=8$. However, $\CT_{\rm rk2}$ is inconsistent, because the gauged $SU(2)$ suffers from Witten's $SU(2)$ anomaly \rcite{Witten:1982fp}.

To understand this last statement, note that the $I_{3,3}$ theory cannot have such an anomaly. Indeed, as we described above, the $I_{3,3}$ theory can be obtained as the IR endpoint of an RG flow from the asymptotically free limit of $SU(2)$ SQCD with $N_f=3$ \rcite{Argyres:1995xn} (the short-distance limit clearly has vanishing Witten anomaly since we can give $SU(2)\subset SO(6)$-preserving masses to the squarks). This flow preserves an $SU(3)\subset SO(6)$ flavor symmetry of the gauge theory, and, moreover, this symmetry is identified with the flavor symmetry of the $I_{3,3}$ theory in the deep IR. Since the RG flow does not leave any additional massless matter besides the $I_{3,3}$ theory at long distances, it must be the case that the Witten anomaly for the $I_{3,3}$ theory matches the (vanishing) Witten anomaly for the UV theory. Therefore, $\CT_{\rm rk2}$ has the same Witten anomaly as the five half-hypermultiplet doublets. This anomaly is clearly non-vanishing, and so the $\CT_{\rm rk2}$ theory is inconsistent. On the other hand, since our $\CT_{2, {3\over2}, {3\over2}}$ theory has an even number of hypermultiplet doublets, it is a consistent theory.

\subsec{Evidence that $\CT_{2,{3\over2},{3\over2}}=I_{4,4}$, and a check of potential cusps}
Let us check that the invariant quantities in \Trkiiidata\ for the $\CT_{2,{3\over2}, {3\over2}}$ theory match the corresponding quantities for the $I_{4,4}$ theory (evidence for the equivalence of the BPS spectra of these theories was given in \rcite{Cecotti:2011rv}). To that end, we first note that, as desired, the $I_{4,4}$ SCFT has the following $\CN=2$ chiral spectrum \rcite{Xie:2012hs, Xie:2013jc}
\eqn{
\CS_{I_{4,4}}=\left\{2,{3\over2},{3\over2}\right\}=\CS_{\CT_{2,{3\over2},{3\over2}}}~.
}[Coulombspecthree]
As a result, using \ShapTach, we find \rcite{Xie:2013jc}
\eqn{
2a_{I_{4,4}}-c_{I_{4,4}}={7\over4}=2a_{\CT_{2,{3\over2},{3\over2}}}-c_{\CT_{2,{3\over2},{3\over2}}}~.
}[Twoamcthree]

Next, we can write down a good UV description of the three dimensional mirror theory.\foot{By this we mean a theory in which the IR superconformal $R$ symmetry is visible in the UV. More precisely, we have in mind a theory in which the IR superconformal $R$-symmetry (or $R$-symmetries if there are multiple sectors) descends from a symmetry (or symmetries if there are multiple sectors) of the RG flow.} According to \rcite{Xie:2013jc}, this theory is described by a quiver involving four $U(1)$ nodes with a bifundamental between each node and the overall $U(1)$ decoupled.\foot{In the prescription of \rcite{Xie:2012hs, Xie:2013jc}, this statement follows from the fact that the irregular singularity of the corresponding Hitchin system has boundary conditions specified by three $4\times 4$ matrices whose eigenvalues are generically different (and whose degeneracies are therefore in one-to-one correspondence with three Young tableaux of the form $[1,1,1,1]$). Note also that we have written the remaining $U(1)$ factors in certain linear combinations that are convenient for applying the mirror symmetry algorithm in \rcite{deBoer:1996ck}.} Deleting a redundant $U(1)$ factor, we find the theory
\begin{align}\label{globalsymiii}
     \begin{tabular}{| c| c |c | c |}
\hline   & $U(1)_A$ & $U(1)_B$ & $U(1)_C$ \cr\hline \hline 
         $Q_1$ & 1 & 0 & 0 \cr \hline 
              $ Q_2$ & 0 & 1 & 0\cr\hline 
               $Q_3$ & 0 & 0 &  1 \cr\hline 
               $Q_4$ & 1 & -1 & 0\cr\hline 
               $Q_5$ & 0 & 1 & -1 \cr \hline 
        $Q_6$ & -1& 0 & 1 \cr \hline 
      \end{tabular}
\end{align}
As a result, we conclude that ${\rm dim}\CM_{\CC}^{\rm3dm}=3$ and therefore \rcite{Xie:2013jc}
\eqn{
a_{I_{4,4}}-c_{I_{4,4}}=-{1\over8}=a_{\CT_{2,{3\over2},{3\over2}}}-c_{\CT_{2,{3\over2},{3\over2}}}~.
}[amcthree]

Finally, we can check that the flavor symmetries match. One way to do this is to take the mirror transform of the above theory (using the algorithm in \rcite{deBoer:1996ck})
\begin{align}\label{globalsymdual}
     \begin{tabular}{|c| c |c | c |}
\hline   & $U(1)_{\hat A}$ & $U(1)_{\hat B}$ & $U(1)_{\hat C}$ \cr\hline\hline 
      $\hat Q_1$ & 1 & 0 & 0 \cr \hline 
       $\hat Q_2$ & 0 & 1 & 0\cr \hline 
        $\hat Q_3$ & 0 & 0 &  1 \cr\hline 
        $\hat Q_4$ & -1 & 0 & 1\cr \hline 
        $\hat Q_5$ & 1 & -1 & 0 \cr \hline  
        $\hat Q_6$ & 0& 1 & -1 \cr \hline 
    \end{tabular}
\end{align}
We see that this theory has a $U(1)^3$ flavor symmetry, and so
\eqn{
G_{I_{4,4}}=U(1)^3=G_{\CT_{2,{3\over2},{3\over2}}}~.
}
Alternatively, we can find the same result directly in the mirror theory by noting that there are three $U(1)$ Coulomb branch symmetries that shift the three independent dual photons by constants. Any additional symmetries would correspond to currents that sit in monopole multiplets of dimension one \rcite{Gaiotto:2008ak}. However, the monopole multiplets have dimension
\eqn{
\Delta(\overrightarrow{a})={1\over2}\left(|a_1|+|a_1-a_2|+|a_2|+|a_2-a_3|+|a_3|+|a_3-a_1|\right)>1~,
}[monopoleDims]
where $\overrightarrow{a}=(a_1, a_2, a_3)\in{\bf Z}^3$ is a magnetic $U(1)^3$ charge vector. Note also that \monopoleDims\ is consistent with the claim that we have a good description of the IR theory since there are no free (or unitarity bound violating) monopole operators in our microscopic description. These results strongly indicate that $\CT_{2,{3\over2},{3\over2}}=I_{4,4}$. 

Let us now ask about possible $S$-dual descriptions. One possibility is that we have various dual descriptions involving an $SU(2)$ gauge group coupled to two $I_{3,3}$ sectors and a doublet of hypermultiplets. A more exotic possibility would involve a dual description with an $SU(2)$ gauge group coupled to a rank-two theory with Coulomb branch spectrum $\left\{{3\over2},{3\over2}\right\}$. While we cannot prove that this second possibility does not occur without the SW analysis of the next section, we can already see it is unlikely. Indeed, it is reasonable to assume that any of the sectors that emerge at the cusps of the conformal manifold are also of class $\CS$ and can be realized as compactifications of the $(2,0)$ $A_k$ theory (since the parent $I_{4,4}$ theory is in this class). However, there are no rank-two theories with spectrum $\left\{{3\over2},{3\over2}\right\}$ that can be built from the recipes in \cite{Xie:2012hs,Xie:2013jc} (besides two decoupled copies of the $I_{3,3}$ theory). In the next subsection, we will demonstrate that the first option described in this paragraph is indeed realized.

\subsec{Analysis of the SW curve}
We begin by writing down the Seiberg-Witten curve for the $I_{4,4}$ theory
\begin{align}
0 = x^4 + qx^2z^2 + z^4 + c_{30}x^3 + c_{03}z^3 + c_{20}x^2 + c_{11}xz + c_{02}z^2 + c_{10}x + c_{01}z + c_{00}~.
\label{eq:curve-I44}
\end{align}
The Seiberg-Witten 1-form is given by $\lambda = xdz$. Since the mass of a BPS state is given by $\oint \lambda$, the 1-form $\lambda$ has scaling dimension one. This observation fixes the scaling dimensions of $x,z, c_{ij}$  and $q$ as
\begin{align}
[x]=[z]=1/2~, \ \ \  [c_{ij}] = 2-\frac{i+j}{2}~,\ \ \  [q]=0~.
\label{eq:dimensions}
\end{align}
The $c_{ij}$ with $0< [c_{ij}]<1$ correspond to relevant couplings of the theory while those with $[c_{ij}]>1$ are regarded as vevs of Coulomb branch operators. The $c_{ij}$ of dimension one are mass-deformation parameters, and the dimensionless parameter $q$ is interpreted as an exactly marginal coupling of the theory.

In order to make contact with the $\CT_{2,{3\over2},{3\over2}}$ theory discussed above, we should first show that an $SU(2)$ gauge symmetry emerges. To that end, let us turn off all the $c_{ij}$ except for $c_{00}$. The SW curve is given by 
\begin{align}
0 = x^4 + qx^2 z^2 + z^4 + c_{00}~.
\label{eq:limit-to-SU2}
\end{align}
In terms of $y= -i(c_{00})^{\frac{3}{2}}/(qx^2),\, \tilde{x}= i\sqrt{c_{00}}z/(\sqrt{2}x),\,f=1-4/q^2$ and $u=c_{00}/q$, this curve is expressed as
\begin{align}
y^2 = (\tilde{x}^2 -u)^2 -f\tilde{x}^4~,
\label{eq:conformalSU2}
\end{align}
with the 1-form now $\lambda =  u\,d\tilde{x}/y$. The equation \eqref{eq:conformalSU2} is precisely the curve for $SU(2)$ with $N_f=4$ \cite{Seiberg:1994aj, Argyres:1995wt}, where $u$ is the Coulomb branch parameter of dimension 2. The parameter $f$ is related to the exactly marginal gauge coupling $\tau = \frac{\theta}{\pi} + \frac{8\pi i}{g^2}$.\footnote{Without loss of generality, we can take $\sqrt{1-f}={2\over q} = {\theta_2^4 +\theta_1^4\over\theta_2^4-\theta_1^4}$ with $\theta_1 =  \sum_{n\in {\bf Z}}e^{\pi i \tau(n+\frac{1}{2})^2}$  and $\theta_2 = \sum_{n\in {\bf Z}}(-1)^n e^{\pi i \tau n^2}$.} 
The equivalence of \eqref{eq:limit-to-SU2} and \eqref{eq:conformalSU2} suggests that the $I_{4,4}$ curve contains a sector described by a conformal $SU(2)$ vector multiplet.

The above $SU(2)$ gauge theory has cusps at $q=\infty$ and $q=\pm2$ where the curve \eqref{eq:conformalSU2} degenerates and different $S$-dual descriptions of the theory become weakly coupled. We can go between the cusps via the transformations $T: \tau \to \tau +1$ and $S: \tau \to -1/\tau$ \cite{Seiberg:1994aj}. In terms of $q$, these are expressed as
$T: q\to \frac{12-2q}{2+q}\,,\; S: q\to -q$. It turns out that $S$ and $T$ can be extended to the full $I_{4,4}$ curve \eqref{eq:curve-I44}. To that end, first consider
\begin{align}
\tilde S: \qquad q\to -q~,\ \ \  c_{k\ell}\to -e^{\frac{\pi i}{4}(\ell-k)} c_{k\ell}~.
\label{eq:I44-S}
\end{align}
 The equation \eqref{eq:curve-I44} is invariant under this transformation after we perform a one-form-preserving coordinate transformation $z\to e^{-\frac{\pi i}{4}}z$ and $x\to e^{\frac{\pi i}{4}}x$. 
 
Next, consider the $T$ transformation. We first shift $x \to x+c_{30}/(2q-4)$ and $z\to z + c_{03}/(2q-4)$ so that the curve \eqref{eq:curve-I44} is
\begin{align}
0 = x^4 + q x^2z^2 + z^4 + (x^2 + z^2)(\tilde{c}_{30}x + \tilde{c}_{03}z)+ \tilde{c}_{20}x^2 + \tilde{c}_{11}xz + \tilde{c}_{02}z^2 + \tilde{c}_{10}x + \tilde{c}_{01}z + \tilde{c}_{00}~.
\label{eq:I44-2}
\end{align}
This shift keeps $\lambda$ invariant up to an exact term.
While the relation between $c_{k\ell}$ and $\tilde{c}_{k\ell}$ is generically complicated, it reduces to $\tilde{c}_{k\ell} = c_{k\ell}$ when $q\to \infty$. Now consider the following transformation:
\begin{align}
\tilde T: \qquad q\to \frac{12-2q}{2+q}~,\ \ \  \tilde{c}_{k\ell} \to \frac{4}{2+q}\ g(\tilde{c}_{k\ell})~,
\label{eq:I44-T}
\end{align}
where $g$ is a linear map defined by $g(\tilde{c}_{30}) = \frac{1}{\sqrt{2}}(\tilde{c}_{30}+ \tilde{c}_{03}),\; g(\tilde{c}_{03}) = \frac{1}{\sqrt{2}}(\tilde{c}_{03}-\tilde{c}_{30}),\; g(\tilde{c}_{20}) = \frac{1}{2}(\tilde{c}_{20} + \tilde{c}_{11} + \tilde{c}_{02}),\; g(\tilde{c}_{11}) = (\tilde{c}_{02}-\tilde{c}_{20}),\; g(\tilde{c}_{02}) = \frac{1}{2}(\tilde{c}_{20} -\tilde{c}_{11} + \tilde{c}_{02}),\; g(\tilde{c}_{10}) = \frac{1}{\sqrt{2}}(\tilde{c}_{10} + \tilde{c}_{01}),\; g(\tilde{c}_{01}) = \frac{1}{\sqrt{2}}(\tilde{c}_{01}-\tilde{c}_{10})$ and $g(\tilde{c}_{00}) = \tilde{c}_{00}$. The equation \eqref{eq:I44-2} is invariant under this transformation after performing a coordinate transformation $x \to \frac{1}{\sqrt{2}}(x+z)$ and $z\to \frac{1}{\sqrt{2}}(z-x)$, which keeps the 1-form invariant up to an exact term. Hence, the $I_{4,4}$ curve is invariant under the transformations generated by $\tilde S$ and $\tilde T$.

As we will show in the remaining parts of this subsection, the cusps of the conformal $SU(2)$ gauge theory persist in the presence of the fractional dimensional operators, and, at each of the cusps $q=\infty, \pm2$, a weakly coupled $SU(2)$ gauge group couples two $I_{3,3}$ theories and a doublet of hypermultiplets. We go between the cusps via the $\tilde S$ and $\tilde T$ transformations (and we use this freedom to study the cusp at $q=\infty$ and then study the $q=\pm2$ cusps via these symmetries).

Moreover, we see in \eqref{eq:I44-S} and \eqref{eq:I44-T} that these transformations act non-trivially on the various parameters and vevs. Note that the $\tilde S$ and $\tilde T$ transformations take a particularly simple form when acting on the independent physical mass parameters (i.e., the independent residues of the one-form), $m_i$ ($i=1,2,3)$, of the theory
\eqna{
\tilde S:& \qquad m_1\to m_1~, \ \ \  m_2\to m_3~, \ \ \ m_3\to m_2~, \cr\tilde T:&\qquad m_1\to m_2~, \ \ \ m_2\to m_1~, \ \ \ m_3\to m_3~,
}
where the $m_i$ are the independent eigenvalues of the simple poles in the Hitchin field at $z=\infty$.\foot{In particular, we have $M_3={\rm diag}(-m_1-m_2-m_3,m_1,m_2,m_3)$ and $M_1={\rm diag}(a,-a,a^{-1},-a^{-1})$ in \eqref{eq:boundary-general}. Turning off the other parameters of the Hitchin system for simplicity, we find $c_{20}=a^{-1}(m_3-m_2)+a(2m_1+m_2+m_3)$, $c_{11}=(a^2-a^{-2})(m_2+m_3)$, $c_{02}=-a^{-1}(2m_1+m_2+m_3)+a(m_2-m_3)$, and $q=-(a^2+a^{-2})$.} As a result, we see that the duality group acts on the residues via $S_3$.

This situation is somewhat reminiscent of the action of the $SL(2,{\bf Z})$ duality group of the $SU(2)$ $N_f=4$ gauge theory on the mass parameters via triality \cite{Seiberg:1994aj} (although here we only have a $U(1)^3$ flavor symmetry instead of $SO(8)$, and we have a non-trivial action of the duality group on the various non-integer dimension parameters of the theory). Indeed, it would be interesting to make this analogy more precise.\foot{In particular, it would be interesting to determine the duality group and any homomorphisms between this group and the group that acts on the parameters of the theory as in \eqref{eq:I44-S} and \eqref{eq:I44-T}.}

\subsubsection{Cusp at $q = \infty$}

Consider the $I_{4,4}$ curve \eqref{eq:curve-I44} near $q = \infty$. Since one of the coefficients is divergent in this limit, it is not clear whether our parameterization of the curve describes the Coulomb branch in a non-singular fashion. As discussed in the introduction, we should normalize the $c_{ij}$ so that the masses of BPS states are non-trivial functions of these quantities.

Let us first consider the Coulomb branch parameter $c_{10}$ of dimension $\frac{3}{2}$. When all the other deformations of the conformal point are turned off, the curve is given by
$0 = x^4 +qx^2z^2 + z^4 + c_{10}x\,$.
To evaluate the periods of this curve, let us change variables as $(x,z)\to (x,w)$ with $w=z/x$. Neglecting a trivial branch ($x=0$), we find
\begin{align}
x^3 = -\frac{c_{10}}{1+qw^2 + w^4}~,
\label{eq:curve-I44-c10}
\end{align}
The 1-form is $\lambda = \frac{1}{2}x^2dw$ up to exact terms. The curve \eqref{eq:curve-I44-c10} is a triple covering of the $w$-plane with branch points at the roots of $1 + qw^2 + w^4$ and at $w=\infty$. Let us define the roots $w_\pm = \pm \sqrt{\frac{1}{2}(-q+\sqrt{q^2-4})}$. In the limit $q\to\infty$, the 1-cycle with the largest absolute value of the period of the one-form is the one around $w=\infty$ and $w=w_{+}$ (or $w_-$). Its period behaves in the limit as
\begin{align}
\frac{1}{2\pi i}\oint \lambda \sim \kappa \frac{(c_{10})^{\frac{2}{3}}}{q^{\frac{1}{2}}},
\label{eq:period-I44-10}
\end{align}
with a $q$-independent constant $\kappa$. Since \eqref{eq:period-I44-10} vanishes in the limit $q\to \infty$, {\it all the periods are vanishing in the limit $q\to \infty$ if $c_{10}$ is finite.} As a result, our parameterization of the Coulomb branch is singular since all finite values of $c_{10}$ are mapped to the origin of the moduli space (i.e., the scale-invariant point). To parameterize the Coulomb branch correctly near $q\sim \infty$, we should normalize $c_{10}$ as
\begin{align}
c_{10} \to q^{\frac{3}{4}}c_{10}~,
\end{align}
so that the period with the largest absolute value remains finite and non-vanishing in the limit $q\to \infty$. We renormalize all the $c_{ij}$ except for $c_{00}$ in the same way (i.e., we demand that the largest period created by each $c_{ij}\neq 0$ remains finite and non-vanishing in the limit $q\to \infty$).

The only deformation we need to study more carefully is $c_{00}$. When only $c_{00}$ is turned on, the curve is 
the genus one curve \eqref{eq:limit-to-SU2}. With an appropriate choice of two independent 1-cycles $\mathcal{A}$ and $\mathcal{B}$, their periods behave in the limit $q\to \infty$ as

\begin{align}\label{monomass}
\frac{1}{2\pi i}\oint_{\mathcal{A}}\lambda \sim 
\sqrt{\frac{c_{00}}{q}}\;,\qquad
\frac{1}{2\pi i}\oint_{\mathcal{B}}\lambda \sim 
\frac{1}{\pi i}\sqrt{\frac{c_{00}}{q}}\log q\;.
\end{align}
Since the ratio of the two periods is divergent, the curve is pinched in the limit $q\to \infty$. This is the signature of a light W-boson and an infinitely massive monopole. A natural normalization in this case is $c_{00} \to qc_{00}$ so that $\frac{1}{2\pi i}\oint_{\mathcal{A}}\lambda \sim \sqrt{c_{00}}$ and $\frac{1}{2\pi i}\oint_{\mathcal{B}}\lambda \sim \frac{1}{\pi i}\sqrt{c_{00}}\log q$.\footnote{We could also normalize $c_{00}$ as $c_{00} \to q(\log q)^2$ so that $\frac{1}{2\pi i}\oint_{\mathcal{A}} \lambda\sim \sqrt{c_{00}}/\log q$ and $\frac{1}{2\pi i}\oint_{\mathcal{B}}\lambda\sim \frac{1}{\pi i}\sqrt{c_{00}}$. Here we use the traditional normalization in which the period of the pinched cycle is finite and non-vanishing. Note that for $c_{ij}\neq c_{00}$ there is a unique renormalization up to $q$-independent rescaling. The reason for this is that no 1-cycle created by $c_{ij}\neq c_{00}$ is pinched in the limit $q\to \infty$.}

As a result, the curve near $q\sim\infty$ is written as
\begin{align}\label{curverenormalised}
0  &= x^4 + qx^2z^2 +z^4 +  q^{\frac{1}{4}}c_{30}x^3 + q^{\frac{1}{4}}c_{03} z^3+ 
q^{\frac{1}{2}}c_{20}x^2 + qc_{11}xz + q^{\frac{1}{2}}c_{02} z^2+ q^{\frac{3}{4}}c_{10}x\cr&+ q^{\frac{3}{4}}c_{01} z + qc_{00}~.   
\end{align}
Let us now study the behavior of this curve in the limit $q\to \infty$. It turns out that the curve splits into three sectors. 
\begin{itemize}

\item

In the region $|z/x|\sim  1/\sqrt{q}$, the curve is well-described by the new set of variables $\tilde{z} = q^{\frac{1}{4}}z$ and $\tilde{x} = q^{-\frac{1}{4}}x$. In the limit $q\to \infty$, the curve reduces to
\begin{align}
0  = \tilde x^4+ \tilde x^2 \tilde z^2 + c_{30} \tilde x^3 +
c_{20}\tilde x^2 + c_{11}\tilde x \tilde z + c_{10}  \tilde x+
c_{00}~.
\end{align}
By shifting $\tilde z\to\tilde z-c_{11}/(2\tilde x)$, the curve is written as
\begin{align}\label{wzero}
0  = \tilde x^4+ \tilde x^2 \tilde z^2 + c_{30} \tilde x^3 +
c_{20}\tilde x^2  + c_{10}  \tilde x+
\left(c_{00} -\frac{c_{11}^2}{4}\right)~.
\end{align}
 The Seiberg-Witten 1-form is now given by $\lambda = -\tilde z d \tilde x $ up to exact terms.
This is exactly the expression for the SW curve of an $I_{3,3}$ theory (under the identifications $(\tilde x, \tilde z)\sim(z,x)$), as given in Eq.~\eqref{eq:curve-I33}.\foot{The minus sign in the 1-form is absorbed by $U(1)_R$ rotation.}
The parameters $c_{30}, c_{20},c_{10}$ are the relevant coupling of dimension $\frac{1}{2}$, a mass parameter, and the vev of the Coulomb branch operator of dimension $\frac{3}{2}$, respectively. The combination $\sqrt{c_{00} - c_{11}^2/4}$ corresponds to the mass parameter associated with an $SU(2)$ subgroup of the $SU(3)$ flavor symmetry.

\item 

In the region $|z/x|\sim \sqrt{q}$, the curve is well-described by the new variables $\tilde{z}=q^{-\frac{1}{4}}z$ and $\tilde{x} = q^{\frac{1}{4}}x$. The 1-form is now $\lambda = \tilde{x}d\tilde{z}$. After shifting $\tilde x \to \tilde x - c_{11}/(2\tilde z)$, the curve in the limit $q\to \infty$ is written as
\begin{align}
0  = \tilde z^4 +  \tilde x^2\tilde z^2+  c_{03}\tilde z^3 +
c_{02}\tilde z^2 +  c_{01} \tilde z +
\left(c_{00} -\frac{c_{11}^2}{4}\right)~.       
\end{align}
This is again an $I_{3,3}$ curve, but now depends on different parameters. The only parameter shared with the previous $I_{3,3}$ curve is the mass $\sqrt{c_{00}-c_{11}/4}$ associated with an $SU(2)$ subgroup of the flavor symmetry. This result suggests that we have gauged a diagonal $SU(2)\subset SU(3)$ of the two $I_{3,3}$ SCFTs.

\item In the region $|z/x|\sim 1$, the curve in the limit $q\to \infty$ is given by
\begin{align}
0 = x^2z^2 + c_{11}xz + c_{00}\;.
\end{align}
This curve describes the $SU(2)$ superconformal QCD in the weak coupling limit with $c_{11}$ a mass parameter for a fundamental hypermultiplet. We can eliminate this term by shifting $x\to x-c_{11}/(2z)$. The curve after the shift is $0 = x^2z^2 + (c_{00} - c_{11}^2/4)$, which describes the pinched W-boson cycle of the weak-coupling $SU(2)$ curve. The mass of the  W-boson is proportional to $\sqrt{c_{00}-c_{11}^2/4}$.\footnote{The shift of the W-boson mass squared by a hypermultiplet mass squared is a common phenomenon. See for example \cite{Seiberg:1994aj}.} The monopole cycle is overlapping between $|z/x|\sim \sqrt{q}\sim\infty$ and $|z/x|\sim 1/\sqrt{q}\sim 0$; its period is divergent.

\end{itemize}
To recapitulate: the first two sectors describe two $I_{3,3}$ theories while the third sector describes an $SU(2)$ vector multiplet coupled to a fundamental hypermultiplet. The W-boson mass implies that the $SU(2)$ sector is gauging the $SU(2)$ flavor subgroups of the $I_{3,3}$ sectors. Hence, the $I_{4,4}$ curve \eqref{eq:curve-I44} near $q\sim \infty$ describes the Coulomb branch of the weak coupling limit of the $\mathcal{T}_{2,\frac{3}{2},\frac{3}{2}}$ theory defined in the introduction.

\subsubsection{Cusps at $q = \pm 2$}

Let us briefly discuss the other cusps at $q=\pm 2$. Since they are mapped to $q=\infty$ by the symmetry transformations $\tilde S$ and $\tilde T$ described in \eqref{eq:I44-S} and \eqref{eq:I44-T}, the theory again splits into two $I_{3,3}$ theories weakly gauged by an $SU(2)$ vectormultiplet coupled to a fundamental hypermultiplet. From \eqref{eq:I44-S} and \eqref{eq:I44-T}, we can read off the renormalized curve near $q\sim \pm 2$ as
\begin{align}
0  = x^4 + qx^2z^2 + z^4 +  \epsilon^{\frac{3}{4}}c_{30}x^3 + \epsilon^{\frac{3}{4}}c_{03} z^3+ 
\epsilon^{\frac{1}{2}}c_{20}x^2 + c_{11}xz + \epsilon^{\frac{1}{2}}c_{02} z^2+ \epsilon^{\frac{1}{4}}c_{10}x  + \epsilon^{\frac{1}{4}}c_{01} z + c_{00}\;,
\end{align}
where $\epsilon = q\mp 2$. It is straightforward to show that, in the limit $q\to\pm2$, the curve splits into two $I_{3,3}$ curves connected by an $SU(2)$ curve. A difference from the previous cusp is that the parameters $c_{ij}$ are now mixed among the three sectors. In terms of the linear map $g$ defined below \eqref{eq:I44-T}, one of the $I_{3,3}$ curves is characterized by $g(c_{30}), g(c_{20}),g(c_{10})$ and $g(c_{00})-g(c_{11})^2/4$ while the other is governed by $g(c_{03}), g(c_{02}),g(c_{01})$ and $g(c_{00})-g(c_{11})^2/4$. The $SU(2)$ vector multiplet and a fundamental hypermultiplet are characterized by $g(c_{00})$ and $g(c_{11})$.

\subsec{The linear quiver}\label{sec2}

In this section, we would like to demonstrate how the $\CT_{2,{3\over2},{3\over2}}$ theory can be engineered from a UV-complete linear quiver. To that end, consider the theory in Figure \ref{fig:quiver1}.
\begin{figure}
\begin{center}
\vskip .5cm
\begin{tikzpicture}[place/.style={circle,draw=blue!50,fill=blue!20,thick,inner sep=0pt,minimum size=6mm},transition/.style={rectangle,draw=black!50,fill=black!20,thick,inner sep=0pt,minimum size=5mm},auto]
\node[transition] (0) at (0,0) [label=below:\textcolor{red}{}] {$1$};
\node[place] (1) at (1.5,0) [shape=circle,label=below:\textcolor{red}{}] {$2$} edge [-] node[auto]{} (0);
\node[place] (2) at (3,0) [shape=circle] {$3$} edge [-] node[auto]{} (1);
\node[place] (3) at (4.5,0) [shape=circle] {$2$} edge [-] node[auto]{} (2);
\node[transition] (7) at (6,0) {$1$} edge[-] (3);
\node[transition] (9) at (3,-1.5) {$1$} edge[-] (2);
\end{tikzpicture}
\caption{A UV linear quiver embedding of the $\CT_{2,{3\over2},{3\over2}}$ theory.}
\label{fig:quiver1}
\end{center}
\end{figure}
Following \cite{Witten:1997sc}, we can write the 
corresponding SW curve as follows: 
\begin{equation}\begin{aligned}& q_1t^2(v+m_1)+t(v^2+\mu_1v+\tilde{u}_2)+(v^3+u_2v+u_3)+\frac{v^2+\mu_2v+u'_2}{t}\Lambda(v+m_2)\\
& +\frac{v+m_3}{t^2}q_2\Lambda^2(v+m_2)^2=0~.\end{aligned}\end{equation} 
The SW differential has the form $\lambda=\frac{v}{t}dt.$ In the above formula $u_i$, $u'_2$ and $\tilde{u}_2$ are the Coulomb 
branch coordinates of the theory, $m_1$ and $m_3$ encode the mass parameters for the two 
$SU(2)$ doublets, $m_2$ is related to the mass of the fundamental hypermultiplet of $SU(3)$, $\mu_1$ and $\mu_2$ are associated 
with the mass parameters of the bifundamental hypermultiplets.\footnote{Notice that the 
above curve is schematic: the parameters $m_i$ are not the physical masses (i.e., the residues of the SW differential) but are instead 
combinations of the mass parameters and the dynamical scale of the theory.} $q_1$, $q_2$ and $\Lambda$ 
are, respectively, the marginal couplings of the $SU(2)$ gauge groups and the dynamical scale of the $SU(3)$ group. If we send one of the $q_i$ couplings to zero, the curve reduces to that of the linear quiver with the $SU(2)$ group replaced by two hypers 
in the fundamental of $SU(3)$, which is indeed the expected degeneration in the ``ungauging" limit. Setting $q_1=q_2=0$ the 
curve reduces to that of $SU(3)$ SQCD with $N_f=5$. If we send to zero $\Lambda$, thus ungauging $SU(3)$, the quiver breaks into two 
pieces, each describing a scale invariant $SU(2)$ theory. Depending on how we write the curve, in the degeneration limit we are left with 
the curve for one of these two sectors. For example, in the above formula, only the terms proportional to a positive power of $t$ 
remain. We can change description and keep the other sector simply with the redefinition $t\rightarrow t/\Lambda$. 
With a constant shift of $v$, which does not affect the form of the SW differential, and a suitable redefinition of the parameters, 
we can bring the curve to the following form, which is more convenient for our later discussion:
\begin{equation}\label{quiver1}q_1t^2(v+m_1)+t(v^2+\mu_1v+\tilde{u}_2)+(v^3+m_2v^2+u_2v+u_3)+\Lambda\frac{v^3+\mu_2v^2+u'_2v}{t}
+\Lambda^2\frac{v^3+m_3v^2}{t^2}q_2=0~.\end{equation}

We are interested in the origin of the moduli space of this theory (i.e., the point in the moduli space we get by setting all the 
parameters in (\ref{quiver1}) to zero except $q_i$) where the curve reduces to  
\begin{equation}q_1t^2v+v^3+q_2\Lambda^2\frac{v^3}{t^2}=0~,\ \ \ \lambda=\frac{v}{t}dt~.\end{equation}
The resulting curve is singular and, as usually happens in 
$\mathcal{N}=2$ theories, the degeneration of the curve signals the presence of a superconformal fixed point, whose SW curve can 
be extracted starting from (\ref{quiver1}) by taking a suitable scaling limit. First of all we define new variables
\begin{equation}\begin{aligned}\label{ciao}t=\sqrt{\Lambda}z~,\; \mu_1=& \sqrt{\Lambda}a_{1/2}~,\; \mu_2=\sqrt{\Lambda}b_{1/2}~,\; u'_2=\sqrt{\Lambda}u_{3/2}~,\; 
\tilde{u}_2=\sqrt{\Lambda}\tilde{u}_{3/2}~,\\
& m_2=\Lambda c_1~,\; u_2=\Lambda m~,\; u_3=\Lambda u~.\end{aligned}\end{equation} 
In terms of these variables, (\ref{quiver1}) becomes 
\begin{equation}\begin{aligned}\Lambda &\left(q_1z^2(v+m_1)+z(\frac{v^2}{\sqrt{\Lambda}}+a_{1/2}v+\tilde{u}_{3/2})+(\frac{v^3}{\Lambda}+c_1v^2+mv+u)\right.\\ 
& \left.+\frac{v^3/\sqrt{\Lambda}+b_{1/2}v^2+u_{3/2}v}{z}+q_2\frac{v^3+m_3v^2}{z^2}\right)=0~.\end{aligned}\end{equation} 
The SW differential is $\lambda=(v/z)dz$. Then, sending $\Lambda$ to infinity, we get the curve
\begin{equation}\label{scft}z^2(v+m_1)+z(va_{1/2}+\tilde{u}_{3/2})-(1+g)(v^2+mv+u)+\frac{b_{1/2}v^2+\tilde u_{3/2}v}{z}+g\frac{v^3+m_3v^2}{z^2}=0~.\end{equation}
To obtain this formula we divided the whole curve by a constant and rescaled $z$ to set to one the coefficient of $z^2v$ and to 
$-1-g$ and $g$ the coefficients of the terms $v^2$ and $v^3/z^2$ respectively. This manipulation is also accompanied by the proper redefinition 
of the parameters. Notice that this transformation does not affect the SW differential. 

Since we are discussing a superconformal theory, all the parameters appearing in (\ref{scft}) should have a definite scaling 
dimension. This can be read from (\ref{ciao}) using the UV dimension of the parameters appearing in (\ref{quiver1}).
Notice that the above curve is homogeneous, in the sense that assigning dimension one to $v$ (which is consistent with the 
constraint on the SW differential) and $1/2$ to $z$ we find that all the terms in (\ref{scft}) have dimension two.
This is precisely the property we expect for the curve describing an SCFT. 

It is straightforward to see that (\ref{scft}) matches the curve describing the $I_{4,4}$ theory in \eqref{eq:curve-I44} (clearly the number and dimensions of the parameters match). To that end, we take $v\to x z$ and make the following transformation which preserves the one-form up to exact terms: $(x, z)\to(Ax+Bz+\kappa_x, Cx+Dz+\kappa_z)$ with $A={i\over\sqrt{2}}{(1+\sqrt{g})^{1\over4}\over(\sqrt{g}-g)^{1\over4}g^{1\over8}}$, $B=A{\sqrt{\sqrt{g}-g}\over g^{1\over4}\sqrt{1+\sqrt{g}}}$, $C=A\sqrt{g}$, $D=-A{\sqrt{\sqrt{g}-g}\over\sqrt{1+\sqrt{g}}}g^{1\over4}$, $\kappa_x={a_{1/2}\over2(1+g)}$, and $\kappa_z={b_{1/2}\over2(1+g)}$. Dividing the resulting equation by $(g-1)/4$ and labeling the coefficients of the various dimensionful terms as in \eqref{eq:curve-I44}, we recover
\eqn{
0 = x^4 + \left({2g+2\over g-1}\right)x^2z^2 + z^4 + c_{30}x^3 + c_{03}z^3 + c_{20}x^2 + c_{11}xz + c_{02}z^2 + c_{10}x + c_{01}z + c_{00}~,
}[resultsubs]
which we recognize as \eqref{eq:curve-I44} with the identification
\eqn{
q={2g+2\over g-1}~.
}[margparrel]

\newsec{A minimal generalization of Argyres and Seiberg's $S$-duality}[AS]

In this section, we turn our attention to the rank four $\CT_{3,2,{3\over2},{3\over2}}$ theory described in the introduction and argue that it exhibits Argyres-Seiberg-like duality (i.e., the quantities {\bf(i)-(iii)} defined in Section~\strategy\ are partitioned differently at the different cusps). In subsection \ref{prelim} we give strong evidence that $\CT_{3,2,{3\over2},{3\over2}}=III_{6,6}^{3\times[2,2,1,1]}$ by matching the invariant quantities {\bf(i)-(iii)}. In subsection \ref{first} we then study the possible partitions of these quantities and find two potential $S$-dual descriptions. 

In subsection \ref{SWcurve} we use the SW curve of the $III_{6,6}^{3\times[2,2,1,1]}$ theory to show that both descriptions we find are indeed realized: at one cusp, we have a perturbative $SU(3)$ gauge group coupled to two $I_{3,3}$ theories and three fundamental flavors (this is our definition of the $\CT_{3,2,{3\over2},{3\over2}}$ theory), while, at another cusp, we find a description with perturbative $SU(2)$ gauge fields coupled to an $I_{3,3}$ theory and an exotic rank two theory we call $\CT_{3,{3\over2}}$ (in the language of \rcite{Xie:2013jc}, this theory can be written as $III_{6,6}^{2\times[2,2,2],[2,2,1,1]}$). One consequence of our study is a derivation of Eq.~\ASkSUii. In the final subsection, we show that $\CT_{3,2,{3\over2},{3\over2}}$ can be embedded in a UV-complete linear quiver.

Before proceeding to the calculations, let us show that---under the same assumptions we used at the beginning of Section~\SW\ to demonstrate the minimality of our first example---there are no rank three theories that exhibit Argyres-Seiberg-like duality.

We can prove this statement as follows. Let us consider the possible rank three theories. They break up into two cases: {\bf(a)} a rank one gauge theory coupled to either a rank two sector or to two rank one sectors, and {\bf(b)} a rank two gauge theory coupled to a rank one sector. Let us consider {\bf(a)} first. In this case, the gauge theory must be $SU(2)$. Let us suppose that it is coupled to two rank one sectors. Vanishing of the one-loop beta function implies that the only possibility is that $SU(2)$ is coupled to two copies of the $I_{3,3}$ theory with an additional doublet. This is the $\CT_{2,{3\over2},{3\over2}}=I_{4,4}$ theory we studied in Section~\SW\ and showed did not exhibit Argyres-Seiberg-like behavior. Next let us suppose that the $SU(2)$ gauge theory is coupled to a rank two sector. In order to have an Argyres-Seiberg-like duality, such a theory must be dual to a rank two gauge theory coupled to a rank one sector with a non-integer dimension Coulomb branch operator as in {\bf(b)}. The possible rank two gauge groups are: $SU(2)\times SU(2)$, $SU(3)$, $Sp(2)$, and $G_2$. We can rule out $Sp(2)$ and $G_2$ immediately since the $I_{2,4}$ and $I_{3,3}$ theories do not have such symmetry groups. The $SU(3)$ case rules out $I_{2,4}$ as well, since it only has $SU(2)$ flavor symmetry. Moreover, the $I_{3,3}$ theory contributes $\delta k_{SU(3)}=3$ to the flavor anomaly. There are no hypermultiplet representations that can contribute $\delta k_{SU(3)}=9$ in order to make the gauging marginal. As a result, we should consider the $SU(2)\times SU(2)$ gauge theory. We cannot use the $I_{2,4}$ theory because it contributes $\delta k_{SU(2)}={8\over3}$ to the flavor anomaly and there are no representations of hypermultiplets that can then make this gauging marginal. On the other hand, if we use the $I_{3,3}$ theory, then we again run into the Witten anomaly we discussed in Section~\SW. Therefore, the simplest generalization of Argyres-Seiberg duality has rank four (like the $\CT_{3,2,{3\over2},{3\over2}}$ theory we are about to study).

\subsec{Preliminary evidence that $\CT_{3,2,{3\over2},{3\over2}}=III_{6,6}^{3\times[2,2,1,1]}$}[prelim]

We will now compute the quantities {\bf(i)-(iii)} described in Section~\strategy\ for the $III_{6,6}^{3\times[2,2,1,1]}$ theory and show that they match the quantities given in \Trkivdata\ for the $\CT_{3,2,{3\over2},{3\over2}}$ SCFT. We will then motivate the existence of an $SU(2)$ gauge theory cusp and demonstrate how these quantities are partitioned at such a point on the conformal manifold.

We first note that the Hitchin system description of $III_{6,6}^{3\times[2,2,1,1]}$ is specified by three $6\times6$ matrices whose eigenvalue degeneracies are encoded in three copies of the Young tableaux $[2,2,1,1]$ (i.e., each matrix has two sets of two-fold degenerate eigenvalues, see Appendix \ref{app:III66}).\foot{This construction is a rank five Hitchin system realization of the theory. It also turns out there is an equivalent rank three realization of the theory: $(III_{4,4}^{3\times[2,1,1]},[2,2])$. This theory has an irregular singularity labeled by the three Young tableaux $[2,1,1]$ (again describing the degeneracy of the eigenvalues of the Hitchin field at the irregular singularity) and a regular singularity labeled by the Young tableau $[2,2]$. Finally, there is also a rank four realization of the theory: $(III_{5,5}^{3\times[2,2,1]},S)$, where the theory has an irregular singularity labeled by the three Young tableaux $[2,2,1]$ combined with a simple regular singularity.} It is straightforward to check that the Coulomb-branch spectrum of the theory is
\eqn{
\CS_{III_{6,6}^{3\times[2,2,1,1]}}=\left\{3,2,{3\over2},{3\over2}\right\}=\CS_{\CT_{3,2,{3\over2},{3\over2}}}~.
}[CoulombSrkiv]
It then follows from \ShapTach\ that
\eqn{
2a_{III_{6,6}^{3\times[2,2,1,1]}}-c_{III_{6,6}^{3\times[2,2,1,1]}}=3=2a_{\CT_{3,2,{3\over2},{3\over2}}}-c_{\CT_{3,2,{3\over2},{3\over2}}}~.
}[Twoamcrkiv]

Next, from the three Young tableaux $[2,2,1,1]$, we use the rules described in \rcite{Xie:2012hs} to write down the three-dimensional mirror
\begin{align}\label{Aside}
     \begin{tabular}{| c | c | c | c | c | c |}
\hline   & $U(1)_1$ &  $U(1)_2$& $U(2)_A$& $U(2)_B$\cr\hline \hline
        $Q_{AB}$ & 0  & 0 &  $2_{+1}$& $2_{-1}$\cr\hline
        $Q_{B1}$ & -1 & 0 &  1& $2_{+1}$\cr\hline
        $Q_{12}$ & +1 &  -1   & 1& 1\cr\hline
        $Q_{2A}$ & 0 & +1 & $2_{-1}$&1\cr\hline
        $Q_{A1}$ & -1 & 0  &$2_{+1}$&1 \cr \hline
        $Q_{B2}$ & 0& -1  &1&$2_{+1}$ \cr\hline
      \end{tabular}
\end{align}
where the subscripts in the $U(2)_{A,B}$ representations are charges under the corresponding $U(1)$ subgroups. Note that the overall $U(1)$ is decoupled and should be eliminated.

As a result, we see that ${\rm dim}\CM_{\CC}^{\rm3dm}=5$, and so
\eqn{
a_{III_{6,6}^{3\times[2,2,1,1]}}-c_{III_{6,6}^{3\times[2,2,1,1]}}=-{5\over24}=a_{\CT_{3,2,{3\over2},{3\over2}}}-c_{\CT_{3,2,{3\over2},{3\over2}}}~.
}[amcRkiv]
To read off the symmetries of the theory, we can apply mirror symmetry again and find
\begin{align}\label{Bside}
     \begin{tabular}{| c | c | c | c |}
\hline   & $U(1)_{\hat A}$ & $U(1)_{\hat B}$ &  $U(2)_C$\cr\hline\hline
        $q_{i=1,2,3}$ & 0 & 0 & $2_{-1}$\cr\hline
        $\hat q$ & 0 & -1 & $2_{+1}$\cr\hline
        $\hat Q$ & -1 & 1 &   1 \cr\hline
        $Q$ & 1 & 0 & $2_{-1}$\cr\hline
      \end{tabular}
\end{align}
This theory has a $U(3)$ flavor symmetry, and so we conclude that
\eqn{
G_{III_{6,6}^{3\times[2,2,1,1]}}=U(3)=G_{\CT_{3,2,{3\over2},{3\over2}}}~.
}[FlavorRkiv]

Alternatively, we can work directly in the mirror theory. Clearly, there is a Coulomb branch symmetry that shifts the three dual photons by independent constants. To see the symmetry enhancement to $U(3)$ in the IR, we should study the monopole operators with dimension one \rcite{Gaiotto:2008ak}. The general formula for the dimensions of the monopole operators is
\eqna{
\Delta(\overrightarrow{a})&={1\over2}\Big(|a_{1,1}|+|a_{A,1}|+|a_{A,2}|+|a_{B,1}|+|a_{B,2}|\Big)+{1\over2}\Big(|a_{1,1}-a_{A,1}|+|a_{1,1}-a_{A,2}|\cr&+|a_{1,1}-a_{B,1}|+|a_{1,1}-a_{B,2}|+|a_{A,1}-a_{B,1}|+|a_{A,1}-a_{B,2}|+|a_{A,2}-a_{B,1}|\cr&+|a_{A,2}-a_{B,2}|\Big)-\Big(|a_{A,1}-a_{A,2}|+|a_{B,1}-a_{B,2}|\Big)~,
}[monopoleDimsRkiv]
where $\overrightarrow{a}=(a_{1,1}, a_{A,1}, a_{A,2}, a_{B,1}, a_{B,2})\in{\bf Z}^5$ is a $U(1)\times U(2)^2$ magnetic charge vector. In writing \monopoleDimsRkiv, we have used the fact that we can shift the magnetic charge by a vector corresponding to the overall decoupled $U(1)$ to set the magnetic flux from the $U(1)_2$ node to zero. It is straightforward to check that, up to unimportant ${\bf Z}_2\times{\bf Z}_2$ permutations, the dimension one monopoles have charges $M_{1}^{\pm}=(0,\pm1,0,0,0)$, $M_2^{\pm}=(0,0,0,\pm1,0)$, $M_3^{\pm}=\pm(0,1,0,1,0)$, which complete the enhancement of $U(1)^3\to U(3)$ in the IR.

\subsec{A first look at the $SU(2)$ cusp and the $\CT_{3,{3\over2}}$ SCFT}[first]

Let us motivate the existence of an $SU(2)$ cusp in the conformal manifold. One way to see such a point should exist is to recall the SW discussion in Section~\SW. Just as we saw the curve for $SU(2)$ with $N_f=4$ emerge when we turned off all the fractional dimensional couplings and vevs, so too we expect the curve for $SU(3)$ with $N_f=6$ to emerge when we turn off the fractional dimensional quantities in the curve of the $\CT_{3,2,{3\over2},{3\over2}}$ theory. From the discussion in \rcite{Argyres:2007cn}, we then expect that there should be a degeneration limit where an $SU(2)$ gauge group emerges. As we will see, the presence of fractional dimensional operators does not spoil this picture, although the emergent sectors that appear are quite different than in \rcite{Argyres:2007cn}.

What can this cusp look like? We again expect a decomposition into sectors of class $\CS$ (of type $A_k$). One possibility is an $SU(2)$ gauge group coupled to a rank one theory with a dimension three Coulomb branch operator, a rank two theory with spectrum $\left\{{3\over2},{3\over2}\right\}$, and some number of fundamentals. However, as we argued in the previous section, such a rank two sector is unlikely to exist in the $A_k$ theories of class $\CS$, and, since our parent theory is of this type, such an option should not be realized. Another possibility is an $SU(2)$ gauge group coupled to a rank three theory with spectrum $\left\{3, {3\over2},{3\over2}\right\}$. However, just as in the case of the rank two theory with spectrum $\left\{{3\over2},{3\over2}\right\}$, such a theory cannot be constructed from the recipes in \rcite{Xie:2012hs, Xie:2013jc}. As a result, the last possibility is an $SU(2)$ gauge group coupled to a rank two theory with Coulomb branch spectrum, $\left\{3,{3\over2}\right\}$, and a copy of the $I_{3,3}$ theory. We denote this rank two theory, $\CT_{3,{3\over2}}$.

Consistency of our picture demands
\eqna{
&a_{\CT_{3,{3\over2}}}=2~, \ \ \ c_{\CT_{3,{3\over2}}}={9\over4}~,\\ & G_{\CT_{3,{3\over2}}}\supset SU(3)\times SU(2)~, \ \ \ k_{SU(3)}=6~, \ \ \ k_{SU(2)}=5~,\\ & \CS_{{\CT_{3,{3\over2}}}}=\left\{3,{3\over2}\right\}~,
}[datarkii]
where the $SU(3)$ flavor symmetry of the $\CT_{3,{3\over2}}$ theory supplies $SU(3)\subset G_{\CT_{3,2,{3\over2},{3\over2}}}$, and the $U(1)\subset G_{\CT_{3,2,{3\over2},{3\over2}}}$ for the total theory is supplied by the $I_{3,3}$ sector.\foot{One nice check of our discussion is the following. If our conjecture is correct, then the fundamental hypermultiplets at the $SU(3)$ cusp are monopoles in the $SU(2)$ gauge theory description (our argument is similar in spirit to the argument in \rcite{Argyres:2007cn}). With this understanding, let us consider $k_{U(1)}$. On the $SU(2)$ gauge theory side of the duality, it is natural to take $k_{U(1)}=k_{SU(3)}^{I_{3,3}}=3$, while on the $SU(3)$ gauge theory side of the duality, $k_{U(1)}=9q^2+9(-q)^2$, with $q$ the $U(1)$ charge of the hypermultiplets. In order to have matching, we need $q={1\over\sqrt{6}}$. Can we show that such a charge suggests that the fundamentals of $SU(3)$ are monopoles of $SU(2)$? Let us consider a monopole state, $|M\rangle$, dressed with $SU(3)$ flavor-singlet fermionic zero modes, $(c, c^{\dagger})$, from the $I_{3,3}$ sector (on the $SU(2)$ gauge theory side of the duality) and $SU(3)$-charged fermionic zero modes from the $\CT_{3,{3\over2}}$ sector, $(d_i, d^{\dagger}_i)$. Note that the $(c, c^{\dagger})$ are charged under the $U(1)$ flavor symmetry while the $(d_i, d_i^{\dagger})$ are not. Indeed, using the fact that the $U(1)\subset SU(3)$ generator of the $I_{3,3}$'s $SU(3)$ flavor symmetry left over after gauging $SU(2)$ is $T={1\over\sqrt{6}}{\rm diag}(1,1,-2)$ (so that ${\rm Tr}\ T^2=1$), we see that the $U(1)$ charge of $c$ is $-{2\over\sqrt{6}}$; we observe there is only one massless hyper left over when we turn on the dimension two vev, since the $I_{3,3}$ theory has a mutually local triplet of charged hypermultiplets. Taking $|M\rangle$ to be an $SU(3)$ singlet with some non-zero $U(1)$ charge, and noting that $\left(\prod d_i^{\dagger}\right)c^{\dagger}|M\rangle$ and $\left(\prod'd_i^{\dagger}\right)|M\rangle$ are CPT conjugates with opposite $U(1)$ charge, we find that the $U(1)$ charge of the monopole state is ${1\over\sqrt{6}}$ as desired.}

We will now argue that the $\CT_{3,{3\over2}}$ theory exists since it can be identified with the following class $\CS$ SCFT
\eqn{
\CT_{3,{3\over2}}=III_{6,6}^{2\times[2,2,2],[2,2,1,1]}~.
}
In other words, we claim that the irregular singularity of the Hitchin system describing this theory has three $6\times6$ matrices with the first two (i.e., those multiplying the third and second order poles at $z=\infty$ in the Higgs field) having three doubly degenerate eigenvalues and the last one (controlling the mass parameters) having two pairs of doubly degenerate eigenvalues.  Indeed, from the discussion in Appendix \ref{app:III66two}, it is straightforward to check that this theory has $\CS_{III_{6,6}^{2\times[2,2,2],[2,2,1,1]}}=\left\{3,{3\over2}\right\}$ and so, from \ShapTach, it has the same $2a-c$ as the $\CT_{3,{3\over2}}$ SCFT.

The three dimensional mirror of the $III_{6,6}^{2\times[2,2,2],[2,2,1,1]}$ theory is somewhat more subtle than the three dimensional mirrors encountered above. From the Hitchin system, we can deduce the following UV description of the three dimensional mirror
\begin{align}\label{Asideii}
     \begin{tabular}{| c | c | c | c | c | c |}
\hline   & $U(1)_1$ &  $U(2)_A$& $U(2)_B$& $U(2)_C$\cr\hline \hline
        $Q_{AB}$ & 0  & $2_{+1}$& $2_{-1}$&1\cr\hline
        $Q_{BC}$ & 0 &  1& $2_{+1}$&$2_{-1}$\cr\hline
        $Q_{CA}$ & 0 &  $2_{-1}$& 1&$2_{+1}$\cr\hline
        $Q_{A1}$  & +1&  $2_{-1}$& 1&1\cr\hline
      \end{tabular}
\end{align}
where the subscripts in the $U(2)_{B,C}$ representations denote charges under the corresponding $U(1)$ subgroups.

Note that all of the nodes in this description are \lq\lq good" in the sense of \rcite{Gaiotto:2008ak}. In particular, the $U(1)_1$ node has $N_f-2N_c=0$ and so too do the $U(2)_{B,C}$ nodes. The $U(2)_A$ node is also good since it has $N_f-2N_c=1$. Therefore, it is natural to guess that this theory should have no monopole operators of dimension $\Delta\le\frac{1}{2}$ and that the flavor symmetry should be $SU(3)\times SU(2)$.\foot{If we regard the theory as an $\CN=1$ theory, then the flavor symmetry would, intriguingly, be $SU(3)\times SU(2)\times U(1)$.} We also find ${\rm dim}\CM_{\CC}^{\rm3dm}=6$ and therefore $a-c=-{1\over4}$.\foot{Note that this value of $a-c$ rules out another potential candidate for describing $\CT_{3,{3\over2}}$: the $III_{6,6}^{3\times[2,2,2]}$ theory.}

This result is certainly compatible with what we expect from \datarkii. However, there is a wrinkle (note that we do not expect the discussion that follows to affect ${\rm dim}\CM_{C}^{\rm 3dm}$ or therefore $a-c$).
Indeed, we can compute the dimensions of the monopole operators \rcite{Gaiotto:2008ak}
\eqna{
\Delta(\overrightarrow{a})=&{1\over2}\Big(|a_{A,1}|+|a_{A,2}|\Big)+{1\over2}\Big(|a_{A,1}-a_{B,1}|+|a_{A,2}-a_{B,1}|+|a_{A,1}-a_{B,2}|\cr&+|a_{A,2}-a_{B,2}|+|a_{B,1}-a_{C,1}|+|a_{B,2}-a_{C,1}|+|a_{B,1}-a_{C,2}|+|a_{B,2}-a_{C,2}|\cr&+|a_{A,1}-a_{C,1}|+|a_{A,2}-a_{C,1}|+|a_{A,1}-a_{C,2}|+|a_{A,2}-a_{C,2}|\Big)\cr&-\Big(|a_{A,1}-a_{A,2}|+|a_{B,1}-a_{B,2}|+|a_{C,1}-a_{C,2}|\Big)~,\cr
}[monopolerkii]
where $\overrightarrow{a}=(a_{A,1}, a_{A,2}, a_{B,1}, a_{B,2}, a_{C,1}, a_{C,2})\in{\bf Z^6}$ is a $U(2)^3$ magnetic charge vector (we have used the freedom of shifting the flux by a charge corresponding to the overall $U(1)$ in order to set the magnetic flux from the $U(1)_1$ node to zero). It is easy to check that, up to ${\bf Z}_2^3$ permutations, the dimension one monopole operators are $M_1^{\pm}=(0,0,\pm1,0,0,0)$, $M_2^{\pm}=(0,0,0,0,\pm1,0)$, $M_3^{\pm}=\pm(0,0,1,0,1,0)$, $M_4^{\pm}=\pm(1,1,1,1,1,1)$, $M_5^{\pm}=\pm(2,0,2,0,2,0)$, $M_6=(1,-1,1,-1,1,-1)$. However, there is also a dimension half monopole operator, $\hat M_{\pm}=\pm(1,0,1,0,1,0)$. The heuristic reason for this result is that the extra $U(2)_A$ we have added to connect the two linear quivers that produce the $SU(3)$ and $SU(2)$ symmetries gives large quantum corrections to the theory. Therefore, even though the quiver is \lq\lq good" by the usual tests, it actually has an apparent dimension half free monopole operator!

If we simply make the assumption that the IR theory consists of a decoupled free multiplet tensored with the remainder of the theory (whose superconformal $R$-symmetry is visible in the UV), then we find that the IR global symmetry group is $SU(2)^2\times SU(3)$ (see also the discussion in \rcite{Cremonesi:2014xha}). In this case, either the extra $SU(2)$ is an accidental symmetry that appears only upon compactifying our four-dimensional theory on $S^1$ and flowing to the IR, or, if it is not, then the $SU(2)$ symmetry that we gauge in the next section should be thought of as a diagonal subgroup of $SU(2)^2$.

\subsec{Analysis of the SW curve}[SWcurve]

The SW curve for the $III_{6,6}^{3\times [2,2,1,1]}$ theory is given by
\begin{align}
0 & =  x^2z^2\left(x + q z\right)\left(x+\frac{z}{q}\right) + b_1x^3z^2 + b_2x^2z^3+ m_1x^3z + m_2xz^3 + m_3x^2z^2
\nonumber\\[1mm]
&\qquad + \left[\left(c_1 + \frac{b_1m_1}{2}\right)x^2z + \left(c_2+ \frac{b_2m_2}{2}\right)xz^2\right]
\nonumber
+ uxz + \frac{m_1^2}{4}x^2 +\frac{m_2^2}{4}z^2\\& \qquad+\frac{m_1c_1}{2}x + \frac{m_2c_2}{2}z + v~,
\label{eq:curve-III66}
\end{align}
with 1-form $\lambda = xdz$. For a derivation of this expression, see Appendix~\ref{app:III66}. The fact that $\oint \lambda$ has scaling dimension one implies $[x]=[z]=\frac{1}{2}$ and
\begin{align}
 [q]=0\,,\quad\; [b_i]=\frac{1}{2}\,,\quad\; [m_a] = 1\,,\quad\; [c_i] = \frac{3}{2}\,,\quad\; [u]=2\,,\quad\; [v]=3\;.
\label{eq:dimensions}
\end{align}
The theory has an exactly marginal coupling $q$ and three mass deformation parameters $m_i$.
The $b_i$ are relevant couplings associated with two Coulomb branch operators of dimension $\frac{3}{2}$, whose vevs are identified with $c_i$.  There are also Coulomb branch operators $u,v$ of integer dimensions.

In order to make contact with the $\CT_{3,2,{3\over2},{3\over2}}$ theory, we should first demonstrate that a conformal $SU(3)$ gauge group emerges in the curve \eqref{eq:curve-III66}. To that end, turning off all the deformations except for $u$ and $v$ yields
\begin{align}
0 &= x^2z^2\left(x+qz\right)\left(x+\frac{z}{q}\right) + uxz + v~.
\label{eq:limit-to-SU3}
\end{align}
In terms of $\tilde{u}= u/[2(q + \frac{1}{q})],\, \tilde{v} = v/[2\sqrt{2}(q+\frac{1}{q})],\,f=4/(q+\frac{1}{q})^2,\,\tilde{x}= xz/\sqrt{2}$ and $y = \tilde{x}^3 + \sqrt{2}z^2\tilde{x}^2/(q+\frac{1}{q}) + \tilde{u}\tilde{x} + \tilde{v}$, the curve is expressed as
\begin{align}
y^2 = (\tilde{x}^3 + \tilde{u}\tilde{x} + \tilde{v})^2 -f\tilde{x}^6~.
\label{eq:conformalSU3}
\end{align}
The 1-form is written as $\lambda = {1\over2\sqrt{2}}\tilde{x}d\log \left(\frac{P-y}{P+y}\right)$ up to exact terms, where $P = \tilde{x}^3 + \tilde{u}\tilde{x} + \tilde{v}$. These are the one-form and curve for the $SU(3)$ gauge theory with $N_f=6$ \cite{Argyres:1995wt}. The parameter $f$ is identified with a modular function of the exactly marginal gauge coupling $\tau = \frac{\theta}{\pi} + \frac{8\pi i}{g^2}$.\footnote{Once again, we take $\sqrt{1-f}= {\theta_2^4 +\theta_1^4\over\theta_2^4-\theta_1^4}$ with $\theta_1 = \sum_{n\in {\bf Z}}e^{\pi i \tau(n+\frac{1}{2})^2}$ and $\theta_2 = \sum_{n\in {\bf Z}}(-1)^n e^{\pi i \tau n^2}$.} The emergence of \eqref{eq:conformalSU3} suggests that the $III_{6,6}^{3\times[2,2,1,1]}$ curve contains a sector described by a conformal $SU(3)$ vector multiplet.

The curve \eqref{eq:conformalSU3} is known to be invariant under $\Gamma(2)\subset SL(2,{\bf Z})$, which is generated by $T^2: \tau \to \tau +2$ and $S: \tau \to -1/\tau$ \cite{Argyres:1995wt}. In terms of $q$, these correspond to $T^2: q\to q\,,\; S: q\to 1/q$. It is clear that the full $III_{6,6}^{3\times [2,2,1,1]}$ curve \eqref{eq:curve-III66} is also invariant under these transformations. Moreover, the curve \eqref{eq:curve-III66} is invariant under 
\begin{align}
q\to -q,\ b_1 \to ib_1,\ b_2\to -ib_2,\  c_1\to ic_1,\  c_2\to -ic_2,  \ u\to -u, \  v\to -v~, \ m_3\to-m_3~,
\label{eq:q_to_-q}
\end{align}
as long as we also send $x\to ix,\, z \to -iz$ (which keeps the 1-form invariant). The two transformations $q\to 1/q$ and $q\to -q$ will be important later in this subsection.
 
The $SU(3)$ superconformal QCD described by \eqref{eq:conformalSU3} has a weak-coupling cusp at $\tau = i\infty$ and a strong coupling cusp $\tau =1$. In terms of $q$, these correspond to $q=0,\infty$ and $q=\pm 1$, respectively. Below we study the behavior of the full $III_{6,6}^{3\times[2,2,1,1]}$ curve \eqref{eq:curve-III66} near these points in the marginal coupling space.

\subsubsection{Cusp at $q = 0,\infty$}

Let us first study the curve near $q= 0,\infty$. Since $q=0$ and $q=\infty$ are related by $q\to 1/q$, we have exactly the same physics at these points. Without loss of generality, we may therefore focus on $q=0$.

We first renormalize all the deformations of the curve so that the largest period created by each deformation is finite and non-vanishing in the limit $q\to 0$. The renormalized curve is written as
\begin{align}
0 & =  x^2z^2\left(x + q z\right)\left(x+\frac{z}{q}\right) + q^{-\frac{1}{2}}\left(b_1x^3z^2 + b_2x^2z^3\right)+ m_1x^3z + m_2xz^3 + q^{-1}m_3x^2z^2
\nonumber\\[1mm]
&\qquad + q^{-\frac{1}{2}}\left[\left(c_1 + \frac{b_1m_1}{2}\right)x^2z + \left(c_2+ \frac{b_2m_2}{2}\right)xz^2\right]
\nonumber
+ q^{-1}uxz + \frac{m_1^2}{4}x^2 +\frac{m_2^2}{4}z^2\\&\qquad + q^{-\frac{1}{2}}\left(\frac{m_1c_1}{2}x + \frac{m_2c_2}{2}z\right) + q^{-1}v~,
\end{align}
which turns out to split into three sectors as follows.
\begin{itemize}
\item In the region $|z/x|\sim q$, we define $\tilde{z} = q^{-\frac{1}{2}}z$ and $\tilde{x} = q^{\frac{1}{2}}x$ so that $|\tilde{z}/\tilde{x}|\sim 1$. In terms of $\tilde{x}$ and $\tilde{z}$, the curve in the limit $q\to 0$ is written as
\begin{align}
0 & =  \tilde{x}^3\tilde{z}^2\left(\tilde{x}+\tilde{z}\right) +  b_1\tilde{x}^3\tilde{z}^2 + m_1\tilde{x}^3\tilde{z} +m_3\tilde{x}^2\tilde{z}^2
\nonumber
+ \left(c_1+ \frac{b_1m_1}{2}\right)\tilde{x}^2\tilde{z} + \left(u\tilde x\tilde z+\frac{m_1^2}{4}\tilde{x}^2\right)\\&\qquad+ \frac{m_1c_1}{2}\tilde{x} + v~,
\end{align}
and the 1-form is given by $\lambda = \tilde{x}d\tilde{z}$ up to exact terms.  Let us shift $\tilde{z}\to\tilde z-\frac{1}{3}(\tilde x+b_1+m_3/\tilde{x})$. This curve can be identified with that of the $(III_{3,3}^{3\times [2,1]},F)$ theory, as given in Eq.~\eqref{eq:curve-III33F} of the appendix, with $m = \frac{1}{2}(m_1-{2m_3\over 3}),\, c_{1/2}=b_1,\, u_{3/2} = c_1-{b_1m_3\over 3},\,\hat{u}=u-\frac{m_3^2}{3}$ and $\hat{v} = v-\frac{m_3u}{3}+\frac{2m_3^3}{27}$. This means that the sector near $|z/x|\sim q$ describes the Coulomb branch of the $(III_{3,3}^{3\times [2,1]},F)$ theory.
 In particular, $\hat{u}$ and $\hat{v}$ are identified with the mass parameters associated with an $SU(3)$ flavor subgroup.

\item In the region $|z/x|\sim 1/q$, we define $\tilde{z} = q^{\frac{1}{2}}z$ and $\tilde{x} = q^{-\frac{1}{2}}x$. The curve in the limit $q\to 0$ is now written as
\begin{align}
0 &=  \tilde{x}^2\tilde{z}^3\left(\tilde{x}+\tilde{z}\right) +  b_2\tilde{x}^2\tilde{z}^3 + m_2\tilde{x}\tilde{z}^3 +m_3\tilde{x}^2\tilde{z}^2
\nonumber
+ \left(c_2+ \frac{b_2m_2}{2}\right)\tilde{x}\tilde{z}^2 + \left(uxz+\frac{m_2^2}{4}\tilde{z}^2\right)\\&\qquad + \frac{m_2c_2}{2}\tilde{z} + v~.
\end{align}
By shifting $\tilde{x}$, we again have the curve \eqref{eq:curve-III33F} for the $(III_{3,3}^{3\times [2,1]},F)$ theory. However, the curve now depends on a different set of parameters. The only parameters shared with the previous $(III_{3,3}^{3\times [2,1]}, F)$ theory are the masses $\hat{u}$ and $\hat{v}$ associated with the $SU(3)$ subgroup of the flavor symmetry.
\item In the region $|z/x|\sim 1$, the curve is
\begin{align}
x^3z^3 + m_3x^2z^2 + uxz + v=0~,
\end{align}
which describes the weak coupling limit of $SU(3)$ superconformal QCD. There are two independent pinched cycles with vanishing intersection, which is a signature of light $SU(3)$ W-bosons and infinitely massive monopoles. The parameter $m_3$ is identified with the mass of a fundamental hypermultiplet. By the shift $z\to z-m_3/(3x)$, the curve is written as $0=x^3z^3 + \hat{u}xz + \hat{v}$; the W-boson masses are then determined by $\hat{u}$ and $\hat{v}$.
\end{itemize}
Hence, in the limit $q\to 0$, the $III_{6,6}^{3\times [2,2,1,1]}$ curve splits into two $(III_{3,3}^{3\times [2,1]},F)$ sectors and an $SU(3)$ gauge sector coupled to a fundamental hypermultiplet. In particular, the $SU(3)$ W-boson masses are associated with the $SU(3)$ flavor subgroup of each $(III_{3,3}^{3\times [2,1]},F)$ theory. This result suggests that the $SU(3)$ sector gauges a diagonal subgroup of the flavor symmetry of each $(III_{3,3}^{3\times [2,1]},F)$ theory and the fundamental hypermultiplet.
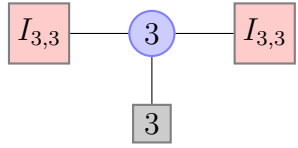
\begin{figure}
\begin{center}
\vskip .5cm
\begin{tikzpicture}[place/.style={circle,draw=blue!50,fill=blue!20,thick,inner sep=0pt,minimum size=6mm},transition/.style={rectangle,draw=black!50,fill=black!20,thick,inner sep=0pt,minimum size=5mm},transition2/.style={rectangle,draw=black!50,fill=red!20,thick,inner sep=0pt,minimum size=8mm},auto]

\node[transition2] (1) at (1.5,0) {$I_{3,3}$};
\node[place] (2) at (3,0) [shape=circle] {$3$} edge [-] node[auto]{} (1);
\node[transition2] (3) at (4.5,0)  {$I_{3,3}$} edge [-] node[auto]{} (2);
\node[transition] (9) at (3,-1.2) {$3$} edge[-] (2);
\end{tikzpicture}
\caption{The quiver diagram describing the $III_{6,6}^{3\times[2,2,1,1]}$ theory at $q\sim 0$. The $U(3)$ flavor symmetry naturally appears from the three fundamental hypermultiplets.} 
\label{fig:quiver1-III66}
\end{center}
\end{figure}

Moreover, as we we explain in Appendix \ref{app:equiv}, each $(III_{3,3}^{3\times[2,1]},F)$ theory can be identified as an $I_{3,3}$ theory with three hypermultiplets. The $I_{3,3}$ theory with three hypermultiplets has $SU(3)_1\times Sp(3)$ flavor symmetry which contains a subgroup $SU(3)_1\times SU(3)_2\times U(1)$. On the other hand, the above discussion shows that the $(III_{3,3}^{3\times [2,1]},F)$ theory has a manifest $SU(3)_3\times U(1)$ flavor symmetry. We identify the $SU(3)_3$ with the diagonal $SU(3)$ of $SU(3)_1\times SU(3)_2$. The three hypermultiplets are then in the fundamental representation of $SU(3)_3$. Since this  $SU(3)_3$ is the one we are gauging in the above discussion, we see that the $III_{6,6}^{3\times [2,2,1,1]}$ theory is identical to the theory $\mathcal{T}_{3,2,\frac{3}{2},\frac{3}{2}}$ of two $I_{3,3}$ theories gauged by a single $SU(3)$ vector multiplet with 3 fundamental hypermultiplets; see Figure~\ref{fig:quiver1-III66}.

\subsubsection{Cusps at $q=\pm 1$}

We now turn to the points $q=\pm1$ in the marginal coupling space. Since these two points are related by the symmetry transformation $q\to -q$, we will, without loss of generality, focus on $q=1$. Note that there is no symmetry transformation which maps $q=\pm 1$ to $q=0,\infty$. Therefore, we expect to have a different weak coupling description in this case.

\begin{figure}
\begin{center}
\vskip .5cm
\begin{tikzpicture}[place/.style={circle,draw=blue!50,fill=blue!20,thick,inner sep=0pt,minimum size=6mm},transition/.style={rectangle,draw=black!50,fill=black!20,thick,inner sep=0pt,minimum size=5mm},transition2/.style={rectangle,draw=black!50,fill=red!20,thick,inner sep=0pt,minimum size=8mm},transition3/.style={rectangle,draw=black!50,fill=red!20,thick,inner sep=0pt,minimum size=10mm},auto]

\node[transition3] (1) at (.4,0) {$\;III_{6,6}^{2\times [2,2,2],[2,2,1,1]}\;$};
\node[place] (2) at (3,0) [shape=circle] {$2$} edge [-] node[auto]{} (1);
\node[transition2] (3) at (4.4,0)  {$I_{3,3}$} edge [-] node[auto]{} (2);
\end{tikzpicture}
\caption{The quiver diagram describing the $III_{6,6}^{3\times [2,2,1,1]}$ theory at $q\sim 1$. The manifest flavor symmetry is $SU(3) \times U(1)$; the $SU(3)$ comes from $\CT_{3,{3\over2}}=III_{6,6}^{2\times [2,2,2],[2,2,1,1]}$, and the $U(1)$ comes from $I_{3,3}$.}
\label{fig:quiver2-III66}
\end{center}
\end{figure}
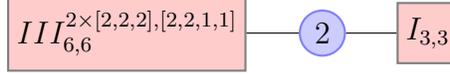

To understand the above statement, we first renormalize the deformations so that the largest period created by each deformation is finite and non-vanishing in the limit $q\to 1$. The correct renormalization turns out to be
\begin{align}
b_1-b_2 \sim \mathcal{O}(\epsilon^{\frac{3}{2}})~,\quad c_1-c_2 \sim \mathcal{O}(\epsilon^{\frac{3}{2}})~,\quad m_3-m_1-m_2-\frac{(b_1+b_2)^2}{16} \sim \mathcal{O}(\epsilon)~,
\end{align}
where $\epsilon = 1-q$.
Therefore the renormalized curve is written as
\begin{align}
0 &=  x^2z^2\left(x + q z\right)\left(x+\frac{z}{q}\right) + x^2z^2\left(\tilde{b}_1 (z+x) + \epsilon^{\frac{3}{2}}\tilde{b}_2(z-x)\right)
\nonumber\\
&\qquad+ xz\left[m_1z(z+x) + m_2x(x+z) + \left(\frac{\tilde{b}_1^2}{4}+\epsilon \tilde{m}_3\right)xz\right]
\nonumber\\[1mm]
&\qquad+ xz\left[\left(\tilde{c}_1 + \frac{(\tilde{b}_1 + \epsilon^{\frac{3}{2}}\tilde{b}_2)m_1}{2}  + \epsilon^{\frac{1}{2}}\tilde{c}_2\right)z + \left(\tilde{c}_1+ \frac{(\tilde{b}_1-\epsilon^{\frac{3}{2}}\tilde{b}_2)m_2}{2}  - \epsilon^{\frac{1}{2}}\tilde{c}_2\right)x\right]
\nonumber\\
&\qquad+ uxz + \frac{m_1^2}{4}z^2 +\frac{m_2^2}{4}x^2 + \frac{m_1(\tilde{c}_1 + \epsilon^{\frac{1}{2}}\tilde{c}_2)}{2}z + \frac{m_2(\tilde{c}_1-\epsilon^{\frac{1}{2}}\tilde{c}_2)}{2}x + v~,
\label{eq:2ndcusp-III66}
\end{align}
where $\tilde{b}_1 = (b_1+b_2)/2,\,\epsilon^{\frac{3}{2}}\tilde{b}_2 = (b_1-b_2)/2,\, \epsilon\tilde{m}_3 = m_3-m_1-m_2-\tilde{b}_1^2/4.$
Let us now define $\zeta = \frac{x+z}{x-z}$. In the limit $q\to 1$, or equivalently $\epsilon \to0$, the curve splits into three sectors, depending on $|\zeta|$.
\begin{itemize}
\item  In the region $|\zeta|\sim 1$, the curve in the limit $q\to 1$ is given by
\begin{align}
0 &=  x^2z^2(x+z)^2 + \tilde{b}_1 x^2z^2(z+x) + m_1xz^2(z+x) + m_2x^2z(x+z) + \frac{\tilde{b}_1^2}{4}x^2z^2
\nonumber\\[1mm]
&\qquad+ xz\left[\left(\tilde{c}_1 + \frac{\tilde{b}_1m_1}{2}\right)z + \left(\tilde{c}_1+ \frac{\tilde{b}_1m_2}{2}\right)x\right]
\nonumber\\
&\qquad+ uxz + \frac{m_1^2}{4}z^2 +\frac{m_2^2}{4}x^2 + \tilde{c}_1\left(\frac{m_1}{2}z + \frac{m_2}{2}x\right) + v~,
\end{align}
This is precisely the curve for the $\text{III}_{6,6}^{2\times [2,2,2],[2,2,1,1]}$ theory, as given in Eq.~\eqref{eq:curve-III66-2} of the appendix. This suggests that the region $|\zeta|\sim 1$ describes the Coulomb branch of the $III_{6,6,}^{2\times [2,2,2],[2,2,1,1]}$ theory.

 \item Now, let us look at the region $|\zeta|\sim \epsilon$. We change variables as $\tilde{x} = \epsilon^{\frac{1}{2}}(x-z)/\sqrt{2i}$ and $\tilde{z}= \epsilon^{-\frac{1}{2}}\sqrt{2i}\left(z +x+ \tilde{b}_1/2 + \frac{m_1-m_2}{x-z}\right)$. It follows that finite $\tilde{x}$ and $\tilde{z}$ correspond to $|\zeta|\sim \epsilon$ in the limit $q\to 1$.
The curve in the limit is now written as
\begin{align}
0 = \tilde{x}^4  + \tilde{z}^2\tilde{x}^2 + \hat{b}_2 \tilde{x}^3 + \hat{m}_3 \tilde{x}^2 + \hat{c}_2\tilde{x} + \hat{u}~,
\label{eq:cuspq1I33}
\end{align}
where $\hat{b}_2 = 2(1-i)\tilde{b}_2,\, \hat{m}_3 = 2i \tilde{m}_3,\, \hat{c}_2 = 4(1+i)\tilde{c}_2,\, \hat{u} = -4u + 2m_1m_2 + 2\tilde{b}_1\tilde{c}_1$. The 1-form is given by $\lambda = \tilde{x}d\tilde{z}$ up to exact terms. Note that this is precisely the curve \eqref{eq:curve-I33} for the $I_{3,3}$ theory. In particular, the mass associated with an $SU(2)$ flavor subgroup is given by $\hat{u}$.

\item  Finally, let us look at the region $|\zeta| \sim \epsilon^{\frac{1}{2}}$, which is between the above two regions. We first define $\tilde{z} = \epsilon^{-\frac{1}{4}}(z+x + \tilde{b}_1/2 + \frac{m_1-m_2}{x-z})$ and $\tilde{x} = \epsilon^{\frac{1}{4}}(x-z)$. It follows that finite $\tilde{x}$ and $\tilde{z}$ correspond to $|\zeta|\sim \epsilon^{\frac{1}{2}}$ in the limit $q\to 1$. The curve in terms of these variables reduces to
\begin{align}
0 = \tilde{x}^2(\tilde{x}^2\tilde{z}^2 + \hat{u})~,
\end{align}
in the limit $q\to 1$.
Apart from the trivial branch $\tilde{x}^2 = 0$, this is the weak coupling limit of the $SU(2)$ curve. The period of the pinched cycle is proportional to $\sqrt{\hat{u}}$, which is identified with the central charge of the $SU(2)$ W-boson.
\end{itemize}
Hence, in the limit $q\to 1$, the $III_{6,6}^{3\times [2,2,1,1]}$ curve splits into three sectors; a $III_{6,6}^{2\times [2,2,2],[2,2,1,1]}$ sector, an $I_{3,3}$ sector, and a perturbative $SU(2)$ gauge sector. This strongly suggests that the $III_{6,6}^{3\times [2,2,1,1]}$ theory at $q\sim 1$ is described by a perturbative $SU(2)$ gauge group coupled to the $\CT_{3,{3\over2}}=III_{6,6}^{2\times [2,2,2],[2,2,1,1]}$ theory and an $I_{3,3}$ theory; see Figure~\ref{fig:quiver2-III66}.

This completes our derivation of the Argyres-Seiberg-like duality in the $\CT_{3,2,{3\over2},{3\over2}}$ SCFT. We see that these results immediately imply that $k_{SU(2)}^{\CT_{3,{3\over2}}}=5$, since the contribution of the $I_{3,3}$ sector to the $SU(2)$ beta function is $k_{SU(2)}^{I_{3,3}}=3$. Moreover, $k_{SU(3)}^{\CT_{3,{3\over2}}}=6$ by anomaly matching. As a result, we have verified equation \ASkSUii\ from the introduction.

\subsec{The linear quiver}
In this subsection, we will show that the $\CT_{3,2,{3\over2},{3\over2}}$ theory can be embedded in a UV-complete linear quiver theory. To understand this claim, let us consider the theory in Figure \ref{fig:quiver2}.
\begin{figure}
\begin{center}
\vskip .5cm
\begin{tikzpicture}[place/.style={circle,draw=blue!50,fill=blue!20,thick,inner sep=0pt,minimum size=6mm},transition/.style={rectangle,draw=black!50,fill=black!20,thick,inner sep=0pt,minimum size=5mm},auto]
\node[place] (1) at (1.5,0) [shape=circle,label=below:\textcolor{red}{}] {$2$};
\node[place] (2) at (3,0) [shape=circle] {$4$} edge [-] node[auto]{} (1);
\node[place] (3) at (4.5,0) [shape=circle] {$2$} edge [-] node[auto]{} (2);
\node[transition] (9) at (3,-1.5) {$3$} edge[-] (2);
\end{tikzpicture}
\caption{A UV linear quiver embedding of the $\CT_{3,2,{3\over2},{3\over2}}$ theory.}
\label{fig:quiver2}
\end{center}
\end{figure}
The SW curve can be written as follows \cite{Witten:1997sc}: 
\begin{equation}\begin{aligned}&q_1t^2(v+m_1)^2+t(v^2+\mu_1v+\tilde{u}_2)(v+m_1)+v^4+u_{02}v^2+u_{03}v+u_{04}+\\ 
&\Lambda\frac{(v^2+\mu_2v+u'_2)(v+m_2)(v+m_3)}{t}+\Lambda^2\frac{q_2(v+m_2)^2(v+m_3)^2}{t^2}=0~.\end{aligned}\end{equation}
The notation is identical to that of Section \ref{sec2}.\footnote{As in Section \ref{sec2}, the above curve is schematic, and 
the parameters $m_i$, $\mu_i$ do not correspond to the physical mass parameters of the theory.} The SW differential is again $\lambda=(v/t)dt$. 
After the shift $v\rightarrow v-m_3$ and a suitable redefinition of the parameters we find the curve 
\begin{equation}\begin{aligned}\label{scc2}&q_1t^2(v+m_1)^2+t(v^2+\mu_1v+\tilde{u}_2)(v+m_1)+v^4-4m_3v^3+u_{02}v^2+u_{03}v+u_{04}+\\ 
&\Lambda\frac{(v^2+\mu_2v+u'_2)(v+m_2)v}{t}+\Lambda^2\frac{q_2(v+m_2)^2v^2}{t^2}=0~.\end{aligned}\end{equation}

By setting all the parameters to zero (apart from $q_1$ and $q_2$) in (\ref{scc2}) the curve becomes singular. Our next task is to 
extract the SW curve describing the effective low-energy theory at this singular point. As in the previous example, we extract the 
curve starting from (\ref{scc2}) and taking a scaling limit. We change variables as follows: 
\begin{equation}\begin{aligned}
& t=\sqrt{\Lambda}z~,\; \mu_1=\sqrt{\Lambda}b_1~,\;\mu_2=\sqrt{\Lambda}b_2~,\;\tilde{u}_2=\sqrt{\Lambda}c_1~,\; 
u'_2=\sqrt{\Lambda}c_2~,\\
& m_3=\Lambda c_0~,\; u_{02}=\Lambda m~,\; u_{03}=\Lambda u_2~,\; u_{04}=\Lambda u_3~.
\end{aligned}\end{equation}
Rewriting (\ref{scc2}) in terms of the new variables and taking the limit $\Lambda\rightarrow\infty$ we find the curve
\begin{equation}\begin{aligned}\label{lin2}&z^2(v+m_1)^2+z(v+m_1)(b_1v+c_1)+(q+\frac{1}{q})v^3+mv^2+u_2v+u_3+\\
&\frac{(v+m_2)v(b_2v+c_2)}{z}+\frac{(v+m_2)^2v^2}{z^2}=0~.\end{aligned}\end{equation}
In the above formula we have divided everything by a constant and rescaled $z$ to set to one the coefficient of the terms 
$z^2v^2$ and $v^4/z^2$. This transformation does not change the SW differential 
$\lambda=(v/z)dz$. We are then left with a single marginal parameter that we call $q$. 

We claim that the above curve describes the theory $III_{6,6}^{3\times[2,2,1,1]}$. Indeed, setting $x=v/z$ we bring the SW differential 
to the canonical form $\lambda=xdz$. The resulting curve is precisely (\ref{eq:curve-III66}) with the identification $u_2=u$ and 
$u_3=v$. The only difference is a factor of two in the definition of $m_1$ and $m_2$.

\newsec{Conclusions}[conclusions]

In this paper, we found minimal generalizations of Seiberg and Witten's $S$-duality in $SU(2)$ gauge theory with four fundamental flavors and Argyres and Seiberg's $S$-duality in $SU(3)$ gauge theory with six fundamental flavors to theories with non-integer dimensional Coulomb branch operators. Along the way, we found an $S$-duality action on the parameters of the $\CT_{2,{3\over2},{3\over2}}$ SCFT that was reminiscent of triality and the emergence of an exotic rank two theory, $\CT_{3,{3\over2}}$, in the case of the $\CT_{3,2,{3\over2},{3\over2}}$ theory.

Many open questions remain. For example, it would be interesting to understand the precise duality group in the case of the $\CT_{2,{3\over2},{3\over2}}$ theory. More generally, it would be interesting to see if we can find new phenomena on the conformal manifolds of $\CN=2$ theories. To study such phenomena, it might be necessary to get a handle on the global properties of these conformal manifolds. For example, two of us recently initiated studies of the global topology of conformal manifolds in certain theories with parametrically small $\CN=2\to\CN=1$ breaking \rcite{Buican:2014sfa}.

At a minimum, it is clear that one can construct many more complicated examples of the $S$-dualities we have discussed in this paper. One promising avenue of investigation would be to see if the type $III$ Hitchin systems we saw naturally arise in our generalization of Argyres-Seiberg duality play an important role in these further studies. We suspect that the IR behavior of the three dimensional theories corresponding to these type $III$ Hitchin systems is typically quite subtle, and it would be useful to develop new tools to analyze the resulting dynamics. Such work might lead to a simple mathematical language that explains the dualities we discussed and points the way to new ones. 

\ack{ \bigskip We would like to thank G. Moore for helpful discussions and comments. M. B. would like to thank the wonderful hospitality and scientific environments of KITP (during the workshop on \lq\lq New Methods in Nonperturbative Quantum Field Theory") and CERN, where parts of this project were completed. T. N. gratefully acknowledges support from the Simons Center for Geometry and Physics, Stony Brook University at which some of the research for this paper was performed. M.B. and T.N. are partly supported by the U.S. Department of Energy under grants DOE-SC0010008, DOE-ARRA-SC0003883, and DOE-DE-SC0007897. This research was supported in part by the National Science Foundation under Grant No. NSF PHY11-25915. S.G. is partially supported by the ERC Advanced Grant ``SyDuGraM", by FNRS-Belgium (convention FRFC PDR T.1025.14 and convention IISN 4.4514.08) and by the ``Communaut\'e Fran\c{c}aise de Belgique" through the ARC program. C.P. is a Royal Society Research Fellow.}

\newpage
\begin{appendix}

\newsec{Hitchin system perspective}[AppA]

In this appendix we briefly review how one obtains the SW curves of various Argyres-Douglas type theories from the corresponding Hitchin system \rcite{Gaiotto:2009hg, Xie:2012hs}. A class of Argyres-Douglas theories are obtained by compactifying the 6d (2,0) theory on a punctured sphere. The Coulomb branch of such a 4d theory (or more precisely its reduction to 3d) is described by the Hitchin system on the sphere with appropriate BPS boundary conditions at the punctures. The $A_{N-1}$ Hitchin system on the punctured sphere involves an $SU(N)$ gauge field and an adjoint $(1,0)$-form $\Phi=\Phi_zdz$ on ${\bf P}^1$, which are constrained by the Hitchin equations
\begin{align}
F + [\Phi,\bar{\Phi}] = 0~,\qquad \bar{\partial}_A\Phi =0~,\qquad \partial_A \bar{\Phi} =0~.
\end{align}
Here $A$ and $F$ are the gauge connection and the curvature, respectively. The differential operator is defined by $\bar{\partial}_A\Phi = (\partial_{\bar{z}}\Phi_z + [A_{\bar{z}},\Phi_z])d\bar{z}\wedge dz$. The Hitchin equations are basically the BPS condition keeping a 4d $\mathcal{N}=2$ supersymmetry. In particular, the Seiberg-Witten curve of the 4d theory is given by the spectral curve
\begin{align}
\det (xdz -\Phi(z)) = 0~,
\end{align}
of the Hitchin system. The spectral curve depends on various parameters of the Hitchin system. Some of them are completely fixed by boundary conditions at the punctures while the others are not. From the 4d viewpoint, the former corresponds to couplings and masses while the latter corresponds to the vevs of Coulomb branch operators.

At the punctures on ${\bf P}^1$, we impose BPS boundary conditions. Since we can trivialize the gauge bundle around the puncture, the boundary condition is given by specifying the singular behavior of $\Phi$ near the puncture. For a trivialized gauge field, $\bar{\partial}_A\Phi=0$ implies $\Phi$ is meromorphic. The singularity at a puncture is called ``regular'' or ``irregular'' if $\Phi$ has a simple or higher-order pole there, respectively. It was shown in \cite{Xie:2012hs} that the resulting 4d theory is an Argyres-Douglas type theory only if there is a single irregular singularity on ${\bf P}^1$ with at most one additional regular singularity. Below, we review the SW curves of several Argyres-Douglas theories of this type.

\subsection{$I_{n,n}$ theory}
\label{app:shift-in-I44}

The $I_{n,n}$ theory is obtained from the $A_{n-1}$ Hitchin system on ${\bf P}^1$ with an irregular singularity.\footnote{Here we use the notation of \cite{Xie:2013jc}. The same theory is called $(A_{n-1},A_{n-1})$ in the language of \cite{Cecotti:2010fi}.} Suppose that the singularity is at $z=\infty$. The boundary condition of the Higgs field $\Phi(z)$ is given by
\begin{align}
\Phi(z) = dz\left[M_1z + M_2 + \frac{M_3}{z} + \mathcal{O}(z^{-2})\right]~,
\label{eq:boundary-general}
\end{align}
where $M_i$ are traceless $n$-by-$n$ matrices. By using gauge transformations, $M_i$ can be simultaneously diagonalized. For the $I_{n,n}$ theory, the matrices $M_i$ can be any diagonal traceless matrices. The lower-order terms of $\mathcal{O}(z^{-2})$ are not fixed by the boundary condition at $z=\infty$ but are subject to the constraint that $\Phi(z)$ is not singular at $z\neq \infty$. The SW curve of the $I_{n,n}$ theory is then given by the spectral curve $\det (xdz-\Phi(z))=0$.

For example, for the $I_{3,3}$ theory, the matrices $M_i$ are arbitrary traceless diagonal $3\times 3$ matrices. Up to coordinate changes keeping $xdz$ invariant, the corresponding spectral curve is written as
\begin{align}
0=z^4 + x^2 z^2 + bz^3 + mz^2 + cz + M^2~,
\label{eq:curve-I33}
\end{align}
where $b,m,M$ are completely fixed by the boundary condition $M_i$, while $c$ is not. Moreover, the fact that $[xdz] = 1$ implies $[b]={1\over 2},\,[m]=[M]=1$ and $[c]={3\over 2}$. This implies that $b$ is a relevant coupling, $m,M$ are mass parameters and $c$ is the vev of a Coulomb branch operator of dimension ${3\over 2}$. The curve \eqref{eq:curve-I33} is indeed identical to the curve for the Argyres-Douglas theory obtained from $SU(2)$ gauge theory with $N_f=3$ flavors \cite{Argyres:1995xn}, which is known to have an $SU(3)$ flavor symmetry. In particular, $M$ is identified with the mass parameter associated with an $SU(2)\subset SU(3)$.

The second non-trivial example is the $I_{4,4}$ theory. The boundary condition \eqref{eq:boundary-general} is now given by $4 \times 4$ matrices $M_i$. Up to coordinate changes, the spectral curve is written as
\begin{align}
0 = x^4 + qx^2 z^2 + z^4 + c_{30}x^3 + c_{03}z^3 + c_{20}x^2 + c_{11}xz + c_{02}z^2 + c_{10}x + c_{01}z + c_{00}~.
\end{align}
Here the dimensions of the parameters are given in \eqref{eq:dimensions}. In particular, this theory has a single exactly marginal coupling, $q$. The spectrum of the Coulomb branch operators is $\{2,\frac{3}{2},\frac{3}{2}\}$.

\subsection{$III_{6,6}^{3\times[2,2,1,1]}$ theory}
\label{app:III66}

Here we consider the $III_{6,6}^{3\times [2,2,1,1]}$ theory, which is obtained from the $A_5$ Hitchin system on ${\bf P}^1$ with an irregular singularity. Suppose that the singularity is at $z=\infty$. The boundary condition for the Higgs field is characterized by \eqref{eq:boundary-general} with three six-by-six matrices $M_i$. In the case of a type $III$ theory, we specify the number of coincident eigenvalues of $M_i$ by Young tableaux \cite{Xie:2012hs}. Since our Young tableaux are now $[2,2,1,1]$, we demand that $M_i$ are of the form
\begin{align}
M_1 &= \text{diag}(\tilde{a}_1,\tilde{a}_1,\tilde{a}_2,\tilde{a}_2,\tilde{a}_3,\tilde{a}_4)~,\cr
M_2 &= \text{diag}(\tilde{b}_1, \tilde{b}_1,\tilde{b}_2,\tilde{b}_2,\tilde{b}_3,\tilde{b}_4)~,\cr
M_3 &= \text{diag}(\tilde{m}_1,\tilde{m}_1,\tilde{m}_2,\tilde{m}_2,\tilde{m}_3,\tilde{m}_4)~,
\end{align}
up to gauge equivalence. Here we implicitly assume the tracelessness of the matrices. This constraint reduces the number of couplings and masses of the corresponding 4d theory.

The SW curve for the $III_{6,6}^{3\times [2,2,1,1]}$ theory is then read off from the spectral curve of the Hitchin system. Up to coordinate changes which keep the 1-form $xdz$ invariant, the spectral curve is written as
\begin{align}
0 & =  x^2z^2\left(x + q z\right)\left(x+\frac{z}{q}\right) + b_1x^3z^2 + b_2x^2z^3+ m_1x^3z + m_2xz^3 + m_3x^2z^2
\nonumber\\[1mm]
&\qquad + \left[\left(c_1 + \frac{b_1m_1}{2}\right)x^2z + \left(c_2+ \frac{b_2m_2}{2}\right)xz^2\right]
\nonumber
+ uxz + \frac{m_1^2}{4}x^2 +\frac{m_2^2}{4}z^2\\& \qquad+\frac{m_1c_1}{2}x + \frac{m_2c_2}{2}z + v~,
\end{align}
where $q,b_i,m_i$ are fixed by the boundary condition while $c_i,u,v$ are not. The fact that $[\oint xdz]=1$ implies $[x]=[z]=1/2$. We then find that $q$ is a marginal coupling, $b_i$ are relevant couplings of dimension $1/2$ and $m_i$ are mass deformation parameters. The $c_i,u$ and $v$ are the vev's of Coulomb branch operators of dimension ${3\over 2},2,3$, respectively.

\subsection{$III_{6,6}^{2\times[2,2,2],[2,2,1,1]}$ theory}
\label{app:III66two}

Let us now consider the $III_{6,6}^{2\times [2,2,2],[2,2,1,1]}$ theory. The Young tableaux imply that the boundary condition at $z=\infty$ is given by
\begin{align}
M_1 &= \text{diag}(\tilde{a}_1,\tilde{a}_1,\tilde{a}_2,\tilde{a}_2,\tilde{a}_3,\tilde{a}_3)~,\cr
M_2 &= \text{diag}(\tilde{b}_1,\tilde{b}_1,\tilde{b}_2,\tilde{b}_2,\tilde{b}_3,\tilde{b}_3)~,\cr
M_3 &= \text{diag}(\tilde{m}_1,\tilde{m}_1,\tilde{m}_2,\tilde{m}_2,\tilde{m}_3,\tilde{m}_4)~,
\end{align}
up to gauge equivalence. We implicitly assume the tracelessness of these matrices. The resulting spectral curve is written as
\begin{align}
0 &= x^2z^2(z+x)^2 + x^2z^2(z+x)b + xz\left(m_1z(x+z) + m_2x(z+x)+ \frac{b^2}{4}xz\right)
\nonumber\\
&\qquad +xz\left[( c+ \frac{bm_1}{2})z + (c+\frac{bm_2}{2})x\right] + \left(\frac{m_1^2}{4}z^2 + \frac{m_2^2}{4}x^2 + uzx\right)
\nonumber\\
&\qquad + \left(\frac{m_1c}{2}z + \frac{m_2c}{2}x\right)+ v~,
\label{eq:curve-III66-2}
\end{align}
up to coordinate changes.
Here $b,m_1,m_2,u$ are fixed by boundary conditions while $c,v$ are not. It follows that $m_1,m_2,\sqrt{u}$ are mass parameters and $b$ is a relevant coupling of dimension ${1\over 2}$. The $c$ and $v$ are vev's of Coulomb branch operators of dimension ${3\over 2}$ and $3$.

\subsection{$(III_{3,3}^{3\times [2,1]},F)$ theory}

The $(III_{3,3}^{3\times [2,1]},F)$ theory is obtained from the $A_2$ Hitchin system on ${\bf P}^1$ with an irregular singularity described by $3\times [2,1]$  Young tableaux and a regular full singularity. Suppose that the irregular one is at $z=\infty$ and the regular one at $z=0$. 
The boundary condition at $z=\infty$ is given by \eqref{eq:boundary-general} with $3 \times 3$ matrices
\begin{align}
M_1 = \text{diag}(\tilde{a},\tilde{a},-2\tilde{a})~,\quad M_2 =\text{diag}(\tilde{b},\tilde{b},-2\tilde{b})~,\quad M_3 = \text{diag}(\tilde{m},\tilde{m},-2\tilde{m})~,
\end{align}
up to gauge equivalence. The boundary condition at $z=0$ is given by
\begin{align}
\Phi(z) \sim dz\left[\frac{1}{z}\left(
\begin{array}{ccc}
\tilde{m}_1&& \\
& \tilde{m}_2 &\\
&& -\tilde{m}_1-\tilde{m}_2\\
\end{array}
\right) + \mathcal{O}(z^{0})\right]~.
\end{align}
Up to coordinate changes, the corresponding spectral curve is written as
\begin{align}
0
& = z^3 + z\left[   \frac{\hat{u}}{x^2}+ \frac{c_{1/2} m+u_{3/2}}{x}  + \left(2m-\frac{c_{1/2}^2}{3}\right)- \frac{2c_{1/2}x}{3}-\frac{x^2}{3}\right] + \frac{\hat{v}}{x^3}+ \frac{3mu_{3/2} - c_{1/2}\hat{u}}{3 x^2} 
\nonumber\\[2mm]
& + \frac{3m^2 -c_{1/2}^2m-c_{1/2}u_{3/2} - \hat{u}}{3x} + \left(\frac{2c_{1/2}^3}{27} - c_{1/2}m  - \frac{u_{3/2}}{3}\right) + \frac{2c_{1/2}^2-6m}{9}x  + \frac{2c_{1/2}x^2}{9}   + \frac{2x^3}{27}~.
\label{eq:curve-III33F}
\end{align}
Here $c_{1/2},m$ are fixed by the boundary condition at $z=\infty$ while $\hat{u},\hat{v}$ are fixed by the one at $z=0$. Therefore these are a relevant coupling and masses. The other parameter, $u_{3/2}$, is regarded as the Coulomb branch parameter. The fact $[xdz] = 1$ implies $[c_{1/2}] = {1\over 2},\, [m] =1,\,[\hat{u}]=2,\,[\hat{v}]=3$ and $[u_{3/2}] = {3\over 2}$. In particular, $\hat{u}$ and $\hat{v}$ are mass parameters for the $SU(3)$ flavor subgroup associated with the full (regular) singularity at $z=0$. In Appendix \ref{app:equiv}, we argue that this theory is identical to the $I_{3,3}$ SCFT with three hypermultiplets.

\section{Equivalence of $(III_{3,3}^{3\times[2,1]},F)$ and $I_{3,3}$ with a triplet of hypermultiplets}
\label{app:equiv}

In this appendix, we show that the $(III_{3,3}^{3\times[2,1]},F)$ SCFT is equivalent to the $I_{3,3}$ theory with a triplet of hypermultiplets. In particular, in \ref{B1} we demonstrate this claim at the level of the SW curves, while in \ref{B2} we demonstrate the equivalence at the levels of the $S^1$ reductions.

\subsec{Seiberg-Witten analysis}[B1]

We will now show that the curve for the $I_{3,3}$ SCFT plus a triplet of hypermultiplets agrees with the curve of the $(III_{3,3}^{3\times[2,1]},F)$ theory.

Let us start from $SU(3)$ gauge fields coupled to the $I_{3,3}$ theory. The corresponding curve can be written in the form \cite{Cecotti:2013lda}
\begin{equation}\label{su31}\frac{\Lambda^b}{t}+x^3+ux+v+(u_{3/2}+xc_{1/2})t+t^2=0~,\quad\lambda=x\frac{dt}{t}~.\end{equation}
In the above formula, $u_{3/2}$ and $c_{1/2}$ represent the chiral operator of dimension $3/2$ and the corresponding coupling constant of the 
$I_{3,3}$ theory. $\Lambda$ is the $SU(3)$ dynamical scale, and $b$ is the corresponding beta function coefficient. 
As discussed in \cite{Cecotti:2013lda}, the matter sector is ``localized'' at $t=\infty$, whereas the above curve near $t=0$ looks like 
the curve for $SU(3)$ SYM. Equivalently, we can say that at $t=0$ we have a trivial matter sector. 

Let us consider an $SU(3)$ gauge theory coupled to a fundamental and to the $I_{3,3}$ theory. In order to write down the SW curve, 
we can start from (\ref{su31}) and replace the trivial sector at $t=0$ with a hypermultiplet in the fundamental of 
$SU(3)$. This modification leads to the curve
\begin{equation}\label{su3} \frac{\Lambda^{b-1}(x+m)}{t}+x^3+ux+v+(u_{3/2}+xc_{1/2})t+t^2=0~,\quad\lambda=x\frac{dt}{t}~,\end{equation} 
We can now consider the redefinition $z=t/(x+m)$, which brings the curve to the form 
\begin{equation}\label{kkk}\frac{\Lambda^{b-1}}{z}+x^3+ux+v+(u_{3/2}+xc_{1/2})(x+m)z+(x+m)^2z^2=0~,\quad\lambda=x\frac{dz}{z}~.\end{equation} 
The SW curve describing the $I_{3,3}$ theory plus three hypermultiplets can be identified just by ungauging. The curve in the form 
(\ref{kkk}) is particularly convenient to that end, since the ungauging can be implemented simply 
by setting $\Lambda$ to zero: in this limit $u$ and $v$ become mass parameters, which is exactly what we expect in the 
ungauging limit. We are then left with  
\begin{equation}x^3+ux+v+(u_{3/2}+xc_{1/2})(x+m)z+(x+m)^2z^2=0~,\quad\lambda=x\frac{dz}{z}~.\end{equation} 
In order to bring it to a ``Gaiotto-inspired form'', we now shift the coordinate $x$ in order to eliminate the term proportional 
to $x^2$. This is simply done by the redefinition \begin{equation}x\rightarrow x-\frac{z^2+zc_{1/2}}{3}~,\end{equation} which does not change the SW differential up to 
exact terms which are irrelevant. We can now write the curve in the canonical form 
\begin{equation}\lambda^3+\lambda\phi_2(z)+\phi_3(z)=0~,\end{equation} where 
\begin{equation}\begin{aligned}\phi_2=&\left(\frac{u}{z^2}+\frac{u_{3/2}+c_{1/2}m}{z}+2m-\frac{c_{1/2}^2}{3}-\frac{2c_{1/2}z}{3}-\frac{z^2}{3}\right)(dz)^2~,\\
\phi_3=& \left(\frac{v}{z^3}+\frac{3u_{3/2}m-c_{1/2}u}{3z^2}+\frac{3m^2-c_{1/2}^2m-u-c_{1/2}u_{3/2}}{3z}+a+\right.\\
& \left.\frac{2c_{1/2}^2-6m}{9}z+\frac{2c_{1/2}z^2}{9}+\frac{2z^3}{27}\right)(dz)^3~.\end{aligned}\end{equation} 
In the above formulas we introduced the parameter 
\begin{equation}a\equiv\frac{2c_{1/2}^3}{27}-c_{1/2}m-\frac{u_{3/2}}{3}~.\end{equation} 
We recognize the curve for the $(III_{3,3}^{3\times[2,1]},F)$ theory \cite{Xie:2013jc}.

\subsec{The $S^1$ reduction of the $(III_{3,3}^{3\times[2,1]},F)$ SCFT}[B2]

Here we will briefly study the $S^1$ compactification of the $(III_{3,3}^{3\times[2,1]},F)$ SCFT. We remind the reader that in Section~\ref{SWcurve}, we saw that two copies of the $(III_{3,3}^{3\times[2,1]},F)$ theory emerge at the $SU(3)$ cusp of the $III_{6,6}^{3\times[2,2,1,1]}$ theory. If our identification $\CT_{3,2,{3\over2},{3\over2}}=III_{6,6}^{3\times[2,2,1,1]}$ is correct, then it must be the case that that $(III_{3,3}^{3\times[2,1]},F)$ is equivalent to a copy of the $I_{3,3}$ theory and a triplet of hypermultiplets, and we indeed saw this equivalence demonstrated at the level of the respective Coulomb branches in the previous subsection. We will now demonstrate the equivalence of the compactified three dimensional theories.

The $S^1$ reduction of $(III_{3,3}^{3\times[2,1]},F)$ can be constructed from the recipe in \rcite{Xie:2012hs}
\begin{align}\label{AAside}
     \begin{tabular}{| c | c | c | c | c |}
\hline   & $U(2)_A$ & $U(1)_B$ & $U(2)_C$& $U(1)_1$\cr\hline\hline
        $Q_{AB}$ & $2_{+1}$ & -1 & 0 & 0\cr\hline
        $Q_{BC}$ & 0 & +1 & $2_{-1}$ & 0 \cr\hline
        $Q_{CA}$ & $2_{-1}$ & 0 &  $2_{+1}$ & 0 \cr\hline
        $Q_{A1}$ & $2_{-1}$ & 0 & $0$&+1\cr\hline
      \end{tabular}
\end{align}
In the above table, the subscripts in the representations signify the charges of the fields under the corresponding $U(1)$ subgroups.

From this data, we can immediately compute the dimensions of the monopole operators \rcite{Gaiotto:2008ak}
\eqna{
\Delta(\overrightarrow{a})=&{1\over2}\Big(|a_{A,1}|+|a_{A,2}|\Big)+{1\over2}\Big(|a_{A,1}-a_{B,1}|+|a_{A,2}-a_{B,1}|+|a_{B,1}-a_{C,1}|\cr&+|a_{B,1}-a_{C,2}|+|a_{A,1}-a_{C,1}|+|a_{A,2}-a_{C,1}|+|a_{A,1}-a_{C,2}|+|a_{A,2}-a_{C,2}|\Big)\cr&-\Big(|a_{A,1}-a_{A,2}|+|a_{C,1}-a_{C,2}|\Big)~,
}[monopoleDimssuiiicusp]
where $\overrightarrow{a}=(a_{A,1}, a_{A,2}, a_{B,1}, a_{C,1}, a_{C,2})\in{\bf Z^5}$ is a magnetic flux vector (we have used the invariance of the theory under shifts by a charge vector corresponding to the overall decoupled $U(1)$ to set the magnetic flux of the $U(1)_1$ factor to zero). The dimension half monopole operators are
\eqn{
M_1^{\pm}=\pm(1,0,1,1,0)~, \ \ \ M_2^{\pm}=\pm(1,0,0,1,0)~, \ \ \ M_3^{\pm}=\pm(0,0,0,1,0)~.
}[freemonopoles]
These operators become three free twisted hypermultiplets in the IR and are the three-dimensional incarnations of the decoupled triplet of hypermultiplets we described above. Note that the existence of the $M_3^{\pm}$ dimension half monopole operator follows immediately from the fact that the $U(2)_C$ node is \lq\lq ugly" in the classification of \rcite{Gaiotto:2008ak} (it has $N_f-2N_c=-1$).

To find the remainder of the theory in the IR, we can follow \rcite{Gaiotto:2008ak} and move along the Coulomb branch of the $U(2)_C$ node by taking $\langle\Phi_C\rangle={\rm diag}(v_1,0)$ (where the $\Phi_{C}$ is the adjoint chiral multiplet of $U(2)_C$) and examine the remaining massless theory.\foot{This essentially amounts to performing a Seiberg-like duality on the $U(2)_C$ node. Such dualities have been studied at the level of the $S^3$ partition function in \rcite{Yaakov:2013fza}.} Turning on this vev in the $\CN=4$ superpotential
\eqna{
W&=\tilde Q_{AB}\Phi_AQ_{AB}-{\rm Tr}\ \left(\Phi_{A}\tilde Q_{CA}Q_{CA}\right)-Q_{A1}\Phi_A\tilde Q_{A1}+\Phi_{B}Q_{BC}\tilde Q_{BC}-\Phi_B\tilde Q_{AB}Q_{AB}\cr&+{\rm Tr}\ \left(\Phi_{C}Q_{CA}\tilde Q_{CA}\right)-Q_{BC}\Phi_{C}\tilde Q_{BC}~,
}[superpot]
leaves (besides a decoupled $U(1)$ parameterizing the moduli space of the free $M_3^{\pm}$ theory \rcite{Gaiotto:2008ak}) a massless theory with $U(2)_C\to U(1)_C$ and the following matter multiplets: $(Q_{AB})_a$, $(Q_{BC})^{2}$, $(Q_{CA})_2^a$, $(Q_{A1})^a$ (along with the corresponding hypermultiplet partners; here $a$ is an $SU(2)_A$ index and $2$ is a $U(2)_C$ index).

This operation turns the $U(2)_A$ node ``ugly'', since now $N_f-2N_c=-1$ at this node. Therefore, following \rcite{Gaiotto:2008ak}, we turn on a vev of the form $\langle\Phi_A\rangle={\rm diag}(v_2,0)$ in the reduced theory. This motion on the Coulomb branch leads to a new theory (again, dropping a decoupled $U(1)$ as before) with $U(2)_A\to U(1)_A$ and the following massless matter multiplets: $(Q_{AB})_2$, $(Q_{BC})^2$, $(Q_{CA})_2^2$, $(Q_{A1})^2$ (and hypermultiplet partners). 

This second operation leaves the $U(1)_1$ node ugly and so we can turn on a vev of the form $\langle\Phi_A\rangle=\langle\Phi_B\rangle=\langle\Phi_C\rangle=v_3$. This shift in vacuum gives mass to the $(Q_{A1})^2, (\tilde Q_{A1})_2$ hypermultiplet leaving over the $S^1$ reduction of the $I_{3,3}$ theory as desired.

\section{An alternative derivation of the curves}
\label{AppC}

The fact that (\ref{scft}) and (\ref{lin2}) are the curves associated with the two models we have studied 
could have been guessed without referring to the Hitchin system. We would now like to illustrate a 
technique which can be applied in many cases to extract the SW curve for an $SU(N)$ gauge theory coupled to a number of hypermultiplets 
in the fundamental representation and to two (generically) nonlagrangian sectors with $SU(N)$ global symmetry. This strategy 
works (at least) when the matter sectors are theories of type $D_p(SU(N))$ studied in \cite{Cecotti:2013lda}.\footnote{The 
same method works for theories with an $SO(2N)$ gauge group and $D_p(SO(2N))$ theories.} In what follows we will focus on this class 
of models. In the language of \cite{Xie:2013jc}, $D_p(SU(N))$ theories with $p\geq N$ correspond to $(I_{N,k},F)$ theories with $k=p-N$. 

Let us start by recalling the SW curve for $SU(N)$ SYM coupled to the $D_p(SU(N))$ theory \cite{Cecotti:2013lda}\footnote{\label{resc} 
The parameters $\alpha_1$ and $\alpha_2$ in (\ref{dpn}) are redundant and do not appear in \cite{Cecotti:2013lda}. Indeed, with the 
redefinition $z\rightarrow z(\alpha_1/\alpha_2)^{1/p}$ and then dividing the resulting equation by $\alpha_1$ we can eliminate them. Of course 
this transformation will affect other terms as well, but this can be cured with a redefinition of the corresponding parameters. 
For later convenience, we will keep them anyway.} 
\begin{equation}\label{dpn}\frac{\Lambda^b}{z}+\alpha_1(v^N+u_2v^{N-2}+\dots+u_N)+\dots+\alpha_2z^p=0~,\ \ \ \lambda=\frac{v}{z}dz~.\end{equation} 
Close to $z=0$, the above curve has the same structure as the curve describing SYM theory with group $SU(N)$.
The parameter $\Lambda$ plays the role of the $SU(N)$ dynamical scale, and setting it to zero corresponds to 
ungauging: the $u_i$ parameters, which play the role of $SU(N)$ Coulomb branch coordinates, become mass parameters of the $SU(N)$ global symmetry that appears in the $\Lambda\rightarrow0$ 
limit. We are then left with the SW curve describing the $D_p(SU(N))$ 
theory. 

To incorporate $n$ fundamental hypermultiplets into the above discussion, we deform the above equation as follows 
\begin{equation}\frac{\Lambda^{b-n}\prod_{i=1}^{n}(v+m_i)}{z}+\alpha_1(v^N+u_2v^{N-2}+\dots+u_N)+\dots+\alpha_2z^p=0~,\ \ \ \lambda=\frac{v}{z}dz~,\end{equation}
where $m_i$ are associated with the mass parameters of the fundamental matter fields. With the one-form-preserving change of variable $t=z/\prod_{i=1}^{n}(v+m_i)$, we get the curve 
\begin{equation}\label{addhyp}\frac{\Lambda^{b-n}}{t}+\alpha_1(v^N+u_2v^{N-2}+\dots+u_N)+\dots+\alpha_2t^p\prod_{i=1}^{n}(v+m_i)^p=0~,\ \ \ \lambda=\frac{v}{t}dt~.\end{equation}
The structure of the curve close to $t=0$ is again that of $SU(N)$ SYM theory. We can thus think of the combined matter sector 
$D_p(SU(N))$ plus $n$ fundamentals as ``localized'' at $t=\infty$ (by this we mean the terms in the above curve proportional 
to positive powers of $t$). We can now introduce a second $D_q(SU(N))$ sector by replacing ${\Lambda^{b-n}/t}$ with the curve of 
the $D_q(SU(N))$ theory (written now in terms of $1/t$) \cite{Cecotti:2013lda}:
\begin{equation}\label{general}\frac{\Lambda^{b'}}{t^q}+\dots+\alpha_1(v^N+u_2v^{N-2}+\dots+u_N)+\dots+\alpha_2t^p\prod_{i=1}^{n}(v+m_i)^p=0~,\ \ \ \lambda=\frac{v}{t}dt~.\end{equation}
For generic choices of $p$, $q$ and $n$ this theory is not conformal, so $b'\neq0$. We can then set $\alpha_1=\alpha_2=1$ (see footnote \ref{resc}), 
provided we redefine $\Lambda$ accordingly. When the theory is conformal, which is the case we are interested in, the story is 
different: $\Lambda^{b'}$ is replaced by a constant which we call $\alpha_3$. With a rescaling of $t$, and dividing the resulting 
equation by a constant, we can reabsorb two of the coefficients $\alpha_{1,2,3}$. The third one is physical and cannot be removed: it parametrizes 
the marginal coupling moduli space.

Following this procedure, we can immediately rederive (\ref{scft}) and (\ref{lin2}). 
Let us consider the $SU(3)$ theory first. The $I_{3,3}$ theory coincides with the $D_2(SU(3))$ model. According to the above 
procedure, the curve for an $SU(3)$ theory coupled to three hypermultiplets in the fundamental and two copies of $I_{3,3}$ can be 
written in the form 
\begin{equation}\begin{aligned}& t^2+(b_1v+c_1)t+(q+\frac{1}{q})(v^3+u_2v+u_3)+\frac{(c_2+b_2v)(v+m_1)(v+m)(v+m_3)}{t}+\\
& \frac{(v+m_1)^2(v+m)^2(v+m_3)^2}{t^2}=0~,\ \ \ \lambda=\frac{v}{t}dt~,\end{aligned}\end{equation}
where we have taken $t\to 1/t$ and $p=q=2$, $n=N=3$, $\alpha_2=\alpha_3=1$ and $\alpha_1=q+1/q$ in (\ref{general}) (the change in sign of the one-form can be absorbed by a $U(1)_R$ transformation). After the redefinition $z=t/(v+m_1)$ and the shift $v\rightarrow v-m$ (and a suitable redefinition of the parameters), we find 
precisely (\ref{lin2}). The SW differential is, up to exact terms, $(v/z)dz$ as in (\ref{lin2}).

In order to extract the curve for the model discussed in Section~\ref{SW}, we first notice that $I_{3,3}$ is also equivalent to the 
$D_4(SU(2))$ theory. The curve for $SU(2)$ coupled to $I_{3,3}$ can then be written as (see (\ref{dpn})) 
\begin{equation}\label{ccd}\frac{\Lambda^{5\over2}}{z}+\alpha_1(v^2+u_2)+az+mz^2+cz^3+\alpha_2z^4=0~,\quad\lambda=\frac{v}{z}dz~.\end{equation}
With a shift of $v$ and by rescaling $z$ we can bring it to the form 
\begin{equation}\frac{\Lambda^{5\over2}}{z}+\alpha_1(v^2+u_2)+(u_{3/2}+a_{1/2}v)z+\alpha'_2(m_1+v)z^2=0~,\quad\lambda=\frac{v}{z}dz~.\end{equation} 
Again following the above recipe, we find that the curve associated with an $SU(2)$ gauge theory coupled to two copies of $I_{3,3}$ 
and to a doublet is 
\begin{equation}z^2(v+m_1)+z(u_{3/2}+a_{1/2}v)-(1+g)(v^2+u_2)+\frac{(\tilde{u}_{3/2}+a_{1/2}v)(v+m)}{z}+g\frac{(v+m_3)(v+m)^2}{z^2}=0~.\end{equation}
In the above formula we have chosen $\alpha_3=1$, $\alpha'_2=g$ and $\alpha_1=-1-g$. After the shift $v\rightarrow v-m$ this becomes precisely (\ref{scft}). 

In fact, we can also use (\ref{ccd}) directly. In this case our prescription leads to the curve 
\begin{equation}z^4+c_1z^3+m_1z^2+a_1z+q(v^2+u_2)+a_2\frac{v+m}{z}+m_2\frac{(v+m)^2}{z^2}+c_2\frac{(v+m)^3}{z^3}+\frac{(v+m)^4}{z^4}=0~.\end{equation}
After the shift $v\rightarrow v-m$ and the subsequent redefinition $x={v\over z}$, which brings the SW differential to the canonical form 
$\lambda=xdz$, the curve becomes identical to (\ref{eq:curve-I44}) (which we saw was, in turn, equivalent to (\ref{scft})). 

\end{appendix}

\newpage
\bibliography{chetdocbib}

\begin{thebibliography}{10}
\ifx\href\asklfhas\newcommand{\href}[2]{#2}\fi
\ifx\arxivref\asklfhas\newcommand{\arxivref}[2]{\href{http://arxiv.org/abs/#1}{#2}}\fi
\ifx\doiref\asklfhas\newcommand{\doiref}[2]{\href{http://dx.doi.org/#1}{#2}}\fi
\parskip 0pt
\normalsize

\bibitem{Asnin:2009xx}
V.~Asnin,
\textit{``{On metric geometry of conformal moduli spaces of four-dimensional
  superconformal theories}''},
\doiref{10.1007/JHEP09(2010)012}{JHEP \textbf{1009}, 012 (2010)},
\normalsize{\texttt{\arxivref{0912.2529}{arXiv:0912.2529}}}.

\bibitem{Seiberg:1994aj}
N.~Seiberg \& E.~Witten,
\textit{``{Monopoles, duality and chiral symmetry breaking in N=2
  supersymmetric QCD}''},
\doiref{10.1016/0550-3213(94)90214-3}{Nucl.Phys. \textbf{B431}, 484 (1994)},
\normalsize{\texttt{\arxivref{hep-th/9408099}{hep-th/9408099}}}.

\bibitem{Montonen:1977sn}
C.~Montonen \& D.I. Olive,
\textit{``{Magnetic Monopoles as Gauge Particles?}''},
\doiref{10.1016/0370-2693(77)90076-4}{Phys.Lett. \textbf{B72}, 117 (1977)}.

\bibitem{Goddard:1976qe}
P.~Goddard, J.~Nuyts \& D.I. Olive,
\textit{``{Gauge Theories and Magnetic Charge}''},
\doiref{10.1016/0550-3213(77)90221-8}{Nucl.Phys. \textbf{B125}, 1 (1977)}.

\bibitem{Witten:1978mh}
E.~Witten \& D.I. Olive,
\textit{``{Supersymmetry Algebras That Include Topological Charges}''},
\doiref{10.1016/0370-2693(78)90357-X}{Phys.Lett. \textbf{B78}, 97 (1978)}.

\bibitem{Osborn:1979tq}
H.~Osborn,
\textit{``{Topological Charges for N=4 Supersymmetric Gauge Theories and
  Monopoles of Spin 1}''},
\doiref{10.1016/0370-2693(79)91118-3}{Phys.Lett. \textbf{B83}, 321 (1979)}.

\bibitem{Argyres:2007cn}
P.C. Argyres \& N.~Seiberg,
\textit{``{S-duality in N=2 supersymmetric gauge theories}''},
\doiref{10.1088/1126-6708/2007/12/088}{JHEP \textbf{0712}, 088 (2007)},
\normalsize{\texttt{\arxivref{0711.0054}{arXiv:0711.0054}}}.

\bibitem{Minahan:1996fg}
J.A. Minahan \& D.~Nemeschansky,
\textit{``{An N=2 superconformal fixed point with E(6) global symmetry}''},
\doiref{10.1016/S0550-3213(96)00552-4}{Nucl.Phys. \textbf{B482}, 142 (1996)},
\normalsize{\texttt{\arxivref{hep-th/9608047}{hep-th/9608047}}}.

\bibitem{Gaiotto:2009we}
D.~Gaiotto,
\textit{``{N=2 dualities}''},
\doiref{10.1007/JHEP08(2012)034}{JHEP \textbf{1208}, 034 (2012)},
\normalsize{\texttt{\arxivref{0904.2715}{arXiv:0904.2715}}}.

\bibitem{Argyres:2007tq}
P.C. Argyres \& J.R. Wittig,
\textit{``{Infinite coupling duals of N=2 gauge theories and new rank 1
  superconformal field theories}''},
\doiref{10.1088/1126-6708/2008/01/074}{JHEP \textbf{0801}, 074 (2008)},
\normalsize{\texttt{\arxivref{0712.2028}{arXiv:0712.2028}}}.

\bibitem{Chacaltana:2010ks}
O.~Chacaltana \& J.~Distler,
\textit{``{Tinkertoys for Gaiotto Duality}''},
\doiref{10.1007/JHEP11(2010)099}{JHEP \textbf{1011}, 099 (2010)},
\normalsize{\texttt{\arxivref{1008.5203}{arXiv:1008.5203}}}.

\bibitem{Argyres:1995jj}
P.C. Argyres \& M.R. Douglas,
\textit{``{New phenomena in SU(3) supersymmetric gauge theory}''},
\doiref{10.1016/0550-3213(95)00281-V}{Nucl.Phys. \textbf{B448}, 93 (1995)},
\normalsize{\texttt{\arxivref{hep-th/9505062}{hep-th/9505062}}}.

\bibitem{Argyres:1995xn}
P.C. Argyres, M.R. Plesser, N.~Seiberg \& E.~Witten,
\textit{``{New N=2 superconformal field theories in four-dimensions}''},
\doiref{10.1016/0550-3213(95)00671-0}{Nucl.Phys. \textbf{B461}, 71 (1996)},
\normalsize{\texttt{\arxivref{hep-th/9511154}{hep-th/9511154}}}.

\bibitem{Xie:2013jc}
D.~Xie \& P.~Zhao,
\textit{``{Central charges and RG flow of strongly-coupled N=2 theory}''},
\doiref{10.1007/JHEP03(2013)006}{JHEP \textbf{1303}, 006 (2013)},
\normalsize{\texttt{\arxivref{1301.0210}{arXiv:1301.0210}}}.

\bibitem{Cecotti:2010fi}
S.~Cecotti, A.~Neitzke \& C.~Vafa,
\textit{``{R-Twisting and 4d/2d Correspondences}''},
\normalsize{\texttt{\arxivref{1006.3435}{arXiv:1006.3435}}}.

\bibitem{Kodaira:1963xx}
K.~Kodaira,
Ann.~of~Math \textbf{77}, 563 (1963).

\bibitem{Giacomelli:2013tia}
S.~Giacomelli,
\textit{``{Confinement and duality in supersymmetric gauge theories}''},
\normalsize{\texttt{\arxivref{1309.5299}{arXiv:1309.5299}}}.

\bibitem{Xie:2012hs}
D.~Xie,
\textit{``{General Argyres-Douglas Theory}''},
\doiref{10.1007/JHEP01(2013)100}{JHEP \textbf{1301}, 100 (2013)},
\normalsize{\texttt{\arxivref{1204.2270}{arXiv:1204.2270}}}.

\bibitem{Witten:1982fp}
E.~Witten,
\textit{``{An SU(2) Anomaly}''},
\doiref{10.1016/0370-2693(82)90728-6}{Phys.Lett. \textbf{B117}, 324 (1982)}.

\bibitem{Buican:2014qla}
M.~Buican, T.~Nishinaka \& C.~Papageorgakis,
\textit{``{Constraints on Chiral Operators in N=2 SCFTs}''},
\normalsize{\texttt{\arxivref{1407.2835}{arXiv:1407.2835}}}.

\bibitem{Buican:2013ica}
M.~Buican,
\textit{``{Minimal Distances Between SCFTs}''},
\doiref{10.1007/JHEP01(2014)155}{JHEP \textbf{1401}, 155 (2014)},
\normalsize{\texttt{\arxivref{1311.1276}{arXiv:1311.1276}}}.

\bibitem{Dolan:2002zh}
F.~Dolan \& H.~Osborn,
\textit{``{On short and semi-short representations for four-dimensional
  superconformal symmetry}''},
\doiref{10.1016/S0003-4916(03)00074-5}{Annals~Phys. \textbf{307}, 41 (2003)},
\normalsize{\texttt{\arxivref{hep-th/0209056}{hep-th/0209056}}}.

\bibitem{Papadodimas:2009eu}
K.~Papadodimas,
\textit{``{Topological Anti-Topological Fusion in Four-Dimensional
  Superconformal Field Theories}''},
\doiref{10.1007/JHEP08(2010)118}{JHEP \textbf{1008}, 118 (2010)},
\normalsize{\texttt{\arxivref{0910.4963}{arXiv:0910.4963}}}.

\bibitem{Aharony:2007dj}
O.~Aharony \& Y.~Tachikawa,
\textit{``{A Holographic computation of the central charges of d=4, N=2
  SCFTs}''},
\doiref{10.1088/1126-6708/2008/01/037}{JHEP \textbf{0801}, 037 (2008)},
\normalsize{\texttt{\arxivref{0711.4532}{arXiv:0711.4532}}}.

\bibitem{Cecotti:2011rv}
S.~Cecotti \& C.~Vafa,
\textit{``{Classification of complete N=2 supersymmetric theories in 4
  dimensions}''},
Surveys~in~differential~geometry,~vol \textbf{18},  (2013),
\normalsize{\texttt{\arxivref{1103.5832}{arXiv:1103.5832}}}.

\bibitem{Gaiotto:2012db}
D.~Gaiotto, G.W. Moore \& A.~Neitzke,
\textit{``{Spectral Networks and Snakes}''},
\doiref{10.1007/s00023-013-0238-8}{Annales~Henri~Poincare \textbf{15}, 61
  (2014)},
\normalsize{\texttt{\arxivref{1209.0866}{arXiv:1209.0866}}}.

\bibitem{Gaiotto:2009hg}
D.~Gaiotto, G.W. Moore \& A.~Neitzke,
\textit{``{Wall-crossing, Hitchin Systems, and the WKB Approximation}''},
\normalsize{\texttt{\arxivref{0907.3987}{arXiv:0907.3987}}}.

\bibitem{Shapere:2008zf}
A.D. Shapere \& Y.~Tachikawa,
\textit{``{Central charges of N=2 superconformal field theories in four
  dimensions}''},
\doiref{10.1088/1126-6708/2008/09/109}{JHEP \textbf{0809}, 109 (2008)},
\normalsize{\texttt{\arxivref{0804.1957}{arXiv:0804.1957}}}.

\bibitem{Nanopoulos:2010bv}
D.~Nanopoulos \& D.~Xie,
\textit{``{More Three Dimensional Mirror Pairs}''},
\doiref{10.1007/JHEP05(2011)071}{JHEP \textbf{1105}, 071 (2011)},
\normalsize{\texttt{\arxivref{1011.1911}{arXiv:1011.1911}}}.

\bibitem{Boalch:XXXXxx}
P.~Boalch,
\textit{``{Irregular Connections and Kac-Moody Root Systems}''},
\normalsize{\texttt{\arxivref{0806.1050}{arXiv:0806.1050}}}.

\bibitem{Boalch:YYYYyy}
P.~Boalch,
\textit{``{Hyperkahler manifolds and nonabelian Hodge theory of (irregular)
  curves}''},
\normalsize{\texttt{\arxivref{1203.6607}{arXiv:1203.6607}}}.

\bibitem{Gaiotto:2008ak}
D.~Gaiotto \& E.~Witten,
\textit{``{S-Duality of Boundary Conditions In N=4 Super Yang-Mills Theory}''},
\doiref{10.4310/ATMP.2009.v13.n3.a5}{Adv.Theor.Math.Phys. \textbf{13}, 721
  (2009)},
\normalsize{\texttt{\arxivref{0807.3720}{arXiv:0807.3720}}}.

\bibitem{Zhiboedov:2013opa}
A.~Zhiboedov,
\textit{``{On Conformal Field Theories With Extremal a/c Values}''},
\doiref{10.1007/JHEP04(2014)038}{JHEP \textbf{1404}, 038 (2014)},
\normalsize{\texttt{\arxivref{1304.6075}{arXiv:1304.6075}}}.

\bibitem{deBoer:1996ck}
J.~de~Boer, K.~Hori, H.~Ooguri, Y.~Oz \& Z.~Yin,
\textit{``{Mirror symmetry in three-dimensional theories, SL(2,Z) and D-brane
  moduli spaces}''},
\doiref{10.1016/S0550-3213(97)00115-6}{Nucl.Phys. \textbf{B493}, 148 (1997)},
\normalsize{\texttt{\arxivref{hep-th/9612131}{hep-th/9612131}}}.

\bibitem{Argyres:1995wt}
P.C. Argyres, M.R. Plesser \& A.D. Shapere,
\textit{``{The Coulomb phase of N=2 supersymmetric QCD}''},
\doiref{10.1103/PhysRevLett.75.1699}{Phys.Rev.Lett. \textbf{75}, 1699 (1995)},
\normalsize{\texttt{\arxivref{hep-th/9505100}{hep-th/9505100}}}.

\bibitem{Witten:1997sc}
E.~Witten,
\textit{``{Solutions of four-dimensional field theories via M theory}''},
\doiref{10.1016/S0550-3213(97)00416-1}{Nucl.Phys. \textbf{B500}, 3 (1997)},
\normalsize{\texttt{\arxivref{hep-th/9703166}{hep-th/9703166}}}.

\bibitem{Cremonesi:2014xha}
S.~Cremonesi, G.~Ferlito, A.~Hanany \& N.~Mekareeya,
\textit{``{Coulomb Branch and The Moduli Space of Instantons}''},
\doiref{10.1007/JHEP12(2014)103}{JHEP \textbf{1412}, 103 (2014)},
\normalsize{\texttt{\arxivref{1408.6835}{arXiv:1408.6835}}}.

\bibitem{Buican:2014sfa}
M.~Buican \& T.~Nishinaka,
\textit{``{Compact Conformal Manifolds}''},
\normalsize{\texttt{\arxivref{1410.3006}{arXiv:1410.3006}}}.

\bibitem{Cecotti:2013lda}
S.~Cecotti, M.~Del~Zotto \& S.~Giacomelli,
\textit{``{More on the $\mathcal{N} =$ 2 superconformal systems of type
  $D_p(G)$}''},
\doiref{10.1007/JHEP04(2013)153}{JHEP \textbf{1304}, 153 (2013)},
\normalsize{\texttt{\arxivref{1303.3149}{arXiv:1303.3149}}}.

\bibitem{Yaakov:2013fza}
I.~Yaakov,
\textit{``{Redeeming Bad Theories}''},
\doiref{10.1007/JHEP11(2013)189}{JHEP \textbf{1311}, 189 (2013)},
\normalsize{\texttt{\arxivref{1303.2769}{arXiv:1303.2769}}}.

\end{thebibliography}

\end{document}